\documentclass[aps, pra, reprint, groupedaddress, amsmath, amssymb, longbibliography]{revtex4-1}

\usepackage{CJK}
\usepackage{graphicx} 
\usepackage{dcolumn} 
\usepackage{bm}
\usepackage[pdfstartview=FitH, CJKbookmarks=true, bookmarksnumbered=true, bookmarksopen=true, colorlinks=true, pdfborder=0 0 1, citecolor=blue, linkcolor=blue, linktocpage=true] {hyperref}
\usepackage{epstopdf} 

\begin{document}
\title{Interaction-induced topological bound states and Thouless pumping in a one-dimensional optical lattice}

\author{Ling Lin$^{1,2}$}

\author{Yongguan Ke$^{1,3}$}

\author{Chaohong Lee$^{1,2,3}$}
\altaffiliation{Email: lichaoh2@mail.sysu.edu.cn; chleecn@gmail.com}

\affiliation{$^{1}$Guangdong Provincial Key Laboratory of Quantum Metrology and Sensing $\&$ School of Physics and Astronomy, Sun Yat-Sen University (Zhuhai Campus), Zhuhai 519082, China}
\affiliation{$^{2}$State Key Laboratory of Optoelectronic Materials and Technologies, Sun Yat-Sen University (Guangzhou Campus), Guangzhou 510275, China}
\affiliation{$^{3}$Nonlinear Physics Centre, Research School of Physics, Australian National University, Canberra ACT 2601, Australia}
\date{\today}


\begin{abstract}
We study topological features of interacting spin-$\frac{1}{2}$ particles in one-dimensional state-dependent optical lattices.
Due to the co-translational symmetry, we introduce the center-of-mass Zak phase with the help of center-of-mass momentum.
There appear topological bound states composed by two particles in different spin states via tuning hopping and interaction strengths.
Under symmetric open boundary conditions, topological edge bound-states appear as a result of the non-trivial center-of-mass Zak phase of bound-state band, which is protected by the center-of-mass inversion symmetry.
The interaction plays a crucial role in the appearance of topological bound states and the system becomes completely trivial if the interaction is switched off.
By periodically modulating the hopping and interaction strengths, we show how to implement topological Thouless pumping of bound states, in which the quantized shift of center-of-mass can be described by a non-trivial center-of-mass Chern number.
\end{abstract}
\maketitle

\section{Introduction}
Topological band theory provides a general framework to explain topological features via topological invariants defined with single-particle energy bands.
It can be traced back to the great success in explaining quantum Hall effects (QHE)~\cite{PhysRevLett.49.405}, in which a so-called TKNN invariant (i.e. Chern number) is used to distinguish different phases of matters.
Topological phase transitions happen when topological invariants change.
Later on, Chern number was linked to the quantized charge pumping in one-dimensional (1D) periodically modulated lattices~\cite{PhysRevB.27.6083}, which has the same origin as QHE.
For decades, topological band theory plays a key role in identifying topological states and exploring topological materials~\cite{RevModPhys.82.3045, PhysRevLett.95.226801, PhysRevLett.61.2015, PhysRevLett.96.106802,PhysRevLett.95.226801,bernevig2006quantum}.

Ultracold atomic systems have been proved as a powerful platform for the investigation of non-interacting topological states, i.e. the realization of Thouless pumping~\citep{lohse2016thouless,nakajima2016topological}, Hofstadter model~\citep{PhysRevLett.111.185301} and Haldane model~\citep{jotzu2014experimental}, measurement of Zak phase ~\citep{atala2013}, dynamical characterization of topological phases ~\citep{meier2018observation,PhysRevLett.121.250403}, etc.
The experimental realizations of the spin-orbit couplings in ultracold atomic gases \citep{PhysRevLett.109.095302,lin2011spin,PhysRevLett.109.095301} also bring a new approach to topological states.
Because the interations between atoms can be precisely tuned by Feshbach resonances, there are increasing interests in  the interplay between interaction and topology in ultracold atom systems \citep{PhysRevX.7.031057,RevModPhys.91.015005}.
Recently, it has been reported that the symmetry-protected topological phase is observed with interacting Rydberg atoms~\citep{de2019observation}.

It is challenging to study the topology in a general interacting many-body system, since conventional topological band theory fails as the inter-particle interaction breaks the single-particle translation symmetry.
To overcome this problem, a method based on twisted boundary condition (TBC) has been utilized to analyze the many-body topological effects~\cite{Niu_1984}.
Another alternative method is the generalized topological band theory regarding the center-of-mass (c.m.) momentum~\cite{PhysRevA.95.063630,Qin_2018}.
Due to strong interaction, the bands of the bound states are isolated from the continuum band \cite{PhysRevA.76.023607,jopB.41.16.161002,PhysRevA.80.015601,PhysRevA.81.043609}.
Such bound states have experimentally demonstrated to be stable even if the interaction is repulsive~\citep{winkler2006repulsively}.
In recent years, topological bound states have been found in various systems, such as, Su-Schrieffer-Heeger (SSH) model~\cite{PhysRevA.94.062704, PhysRevA.95.053866, PhysRevB.95.115443, Marques_2018}, XXZ chain~\cite{PhysRevB.96.195134}, Haldane model~\cite{PhysRevA.97.013637}, Hofstadter superlattice model~\cite{Qin_2018}, Rice-Mele model~\cite{PhysRevA.95.063630}, and Floquet system~\cite{zhong2017floquet}.
Among these models, we note that they may support topological states even in the absence of interaction.
A natural question arises: \emph{are there topological states in an interacting multi-particle system whose non-interacting counterpart does not support any topological state?}
Furthermore, due to the interaction, some discrete symmetries essential to topology are reduced, such as chiral symmetry and inversion symmetry.
Therefore, we are wondering \emph{if any essential discrete symmetry is still preserved in 1D interacting systems}.

In this work, we study interacting spin-1/2 particles in a one-dimensional state-dependent lattice, in which the hopping strengthens are state-dependent and interaction strengthens are distance-dependent.
We first calculate the two-particle energy bands with respect to c.m. momentum.
We find the existence of interaction-induced topological bound states which are characterized by c.m. Zak phase.
Remarkably, the topological bound states is protected by the c.m. inversion symmetry, and the c.m. Zak phase is quantized if the interspecies interactions are the same.
The topological edge bound-states under open boundary conditions are supported by nontrivial two-body Zak phases, indicating the existence of bulk-boundary correspondence in the interacting systems.
Moreover, we propose a scheme of implementing topological Thouless pumping via modulating the interactions and tunnelings.
The non-trivial c.m. Chern number indicates a quantized shift of c.m. position, which is verified by our numerical simulation.
We emphasize that both the topological bound state and the topological transport are completely induced by interaction effects.

The paper is organized as follows.
In Sec.~\ref{Sec:General_Theory}, we introduce the c.m. Zak phase and Chern number based upon the c.m. momentum.
In Sec.~\ref{Sec:Model}, we describe our two-particle system in a 1D state-dependent optical lattice.
%
%
In Sec.~\ref{Sec:Solution_BSs}, we calculate the Bloch Hamiltonian and give the energy band structure with respect to the c.m. momentum.
In Sec.~\ref{Sec:Topological_BSs}, we investigate the topology of the isolated bands with the help of c.m. Zak phase.
In Sec.~\ref{Sec:Pumping}, we propose a scheme to implement the interaction-induced topological Thouless pumping.
Finally, we give a summary and discuss our results in Sec.~\ref{Sec:Summary}.

\section{Cotranslational symmetry and center-of-mass Zak phase}
\subsection{General formalism}
\label{Sec:General_Theory}
For interacting system under the periodic boundary condition (PBC), the quasi-momentum of single particle is no more the conserved quantity.
Considering interaction that only depends on relative position between particles, the Hamiltonian
\begin{eqnarray}
H &=& H_T + V \nonumber \\
   &=& \sum\limits_i {{H_{T,i}}}  + \sum\limits_{i < j} {V\left( {|{r_i} - {r_j}|} \right)}
\end{eqnarray}
is invariant under the cotranslation of all $N$ particles through a unit cell in one-dimensional lattice with periodic boundary condition \cite{PhysRevB.96.195134, Qin_2018}.
For simplicity, we consider a normal 1D lattice with $N$ particles and $L$ sites here.
The cotranslation operation is formulated by
\begin{equation}
\hat{\mathcal{T}}|{r_1},{r_2},...,{r_N}\rangle  = |{r_1} + 1,{r_2} + 1,...,{r_N} + 1\rangle ,
\end{equation}
where $\hat{\mathcal{T}}$ is the cotranslation operator, $r_n,\; n=1,2,...,N$ denotes the position of $n$-th particle.
The lattice constant is set to unity $a=1$ by default.
With the cotranslation symmetry, one has $[ {\hat H,\hat{\mathcal{T}}} ] = 0$.
The corresponding conserved quantity is the center-of-mass (c.m.) quasi momentum $K$ of all particles.
Before proceeding further, we introduce the concept of \emph{seed basis} \cite{PhysRevA.95.063630}.
Generally speaking, the seed basis are all the possible states that can \emph{not} be generated to each other by cotranslation operator $\hat{\mathcal{T}}$.
The total number of seed basis depends on the number of particles and the geometry of lattice.
Physically, each element of the seed basis corresponds to a certain distribution of particles, and we shall denote this set as $\{ |{r_1},...,{r_N}\rangle \}$.
The choice of seed basis seems to be somewhat arbitrary.
For example, given a certain particle distribution ${r_1},...,{r_N}$, one can choose either $|{r_1},...,{r_N}\rangle $ or $ |{r_1+d},...,{r_N+d}\rangle, \; d \in \mathbb{Z}$ as one of the seed basis.
In fact, different choices of seed basis correspond to different \emph{gauge} choices for $|K,\alpha\rangle$ \citep{PhysRevB.91.125424}.
Although the choice of gauge do not affect the energy bands, it may affect the calculation of Berry phase.
We would like to remark that different elements of seed basis can be considered as different virtual ``orbits" labelled by $\alpha$.
In this angle, the many-body system can be considered as a single quasiparticle.
Thus, the generalization from band theory of single particle to the many-body case is quite natural and reasonable.
Eigenstates of cotranslation operator $\hat{\mathcal{T}}$ are found to be the superposition of a series states generated by the any of the given seed basis
\begin{eqnarray}
\label{eq:Momentum_Eigenstates}
|K,{\alpha}\rangle  &=&  \frac{1}{\sqrt{L}}\sum\limits_{l=0} ^{L-1}{e^{iK(l + \Sigma_i{r_i}/N)}}\hat{\mathcal{T}}^l |\{ r_{\alpha}\}\rangle  \nonumber \\
&=& \frac{1}{\sqrt{L}}\sum\limits_{l=0} ^{L-1}{{e^{iK(l + \Sigma_i{r_i}/N)}}|\{ r_{\alpha}\} + l\rangle  },
\end{eqnarray}
where $|\{ r_{\alpha}\} + l\rangle \equiv |r_1 + l, r_2 + l, ..., r_N+ l\rangle $, and $\{ r_{\alpha}\}$ denotes one of the given seed basis $|r_1, r_2, ..., r_N\rangle $ which is characterized by $\alpha$.
The eigenvalues of $\hat{\mathcal{T}}$ can be derived as
\begin{eqnarray}
\hat T|K,&&\alpha \rangle  = \sum\limits_l {{e^{iK(l + \Sigma_i{r_i}/N)}}\hat{T}|{r_1} + l,...,{r_N} + l\rangle }  \nonumber \\
&& = \sum\limits_l {{e^{iK(l + \Sigma_i{r_i}/N)}}|{r_1} + (l + 1),...,{r_N} + (l+1)\rangle } 		 \nonumber \\
&& = {e^{ - iK}}\sum\limits_l {{e^{iK(l + 1 + \Sigma_i{r_i}/N)}}|{r_1} + (l + 1),...,({r_N} + l + 1)\rangle } 		\nonumber \\
&& = {e^{ - iK}}|K,{\alpha}\rangle .
\end{eqnarray}
For a lattice with PBC, there is $r_n+L=r_n$, and therefore ${{\hat{\mathcal{T}}}^L}$ is an identity matrix, yielding ${e^{ - iLK}} = 1,K=2\pi n/L, n \in \mathbb{Z}$.
The Hamiltonian can be block-diagonalized as the summation of Bloch Hamiltonians with momentum $K$,
\begin{equation}
\hat H = { \oplus_K}\hat H\left( K \right),
\end{equation}
where
\begin{equation}
{H_{\alpha ',\alpha }}\left( K \right) = \langle K,\alpha '|\hat H|K,{\alpha}\rangle .
\end{equation}
The eigenstates of $\hat{H}(K)$ are the linear combinations of seed basis
\begin{equation}
\label{Def:ManyBody_BlochStates}
|\psi _K^n\rangle  = \sum\limits_{\{\alpha \} } {{u^n_{K,\alpha }}|K,{\alpha }\rangle },
\end{equation}
in which $\{ \alpha \}$ implies the summation of $|K,\alpha\rangle$ with respect to all seed basis labelled by $\alpha$.
By solving the eigenvalue problem of $H(K)$, one can obtain the eigenvectors $u^n_{K,\alpha }$ and eigenvalues $E_K^n$.
The Brillouin Zone (B.Z.) with respect to c.m. momentum $K$ forms a manifold, it is natrual to investigate the topology of this manifold.
Here, with respect to the c.m. momentum, we introduce the concepts of c.m. Zak phase \cite{PhysRevLett.62.2747}
\begin{equation}
\label{Def:ZakPhase}
{\gamma^n _{{\rm{Zak}}}} = i\int_{ - \pi }^\pi  {\langle u_K^n|{\partial _K}|u_K^n\rangle dK} ,
\end{equation}
for 1D system (we mention it as Zak phase for short in the following), and c.m. Chern number \cite{PhysRevA.95.063630}
\begin{equation}
\label{Def:ChernNum}
{C_n} = \frac{i}{{2\pi }}\int_{B.Z.} {dK\int_0^T {dt\left( {\langle {\partial _t}{u_K^n}|{\partial _K}{u _K^n}\rangle  - \langle {\partial _K}{u _K^n}|{\partial _t}{u _K^n}\rangle } \right)} }
\end{equation}
for (1+1)D system where the Hamiltonian is periodically modulated with period $T$.
This is quite similar to the well-known topological band theory in non-interacting system.
It has been shown that the c.m. shift of multi-particle Wannier state is related to the c.m. Chern number \cite{PhysRevA.95.063630} in the topological pumping process.
It is tempting to investigate the physical interpretation of c.m. Zak phase.
According to the modern theory of polarization, it is known that the Zak phase in the non-interacting case is related to the Wannier center (or band center \cite{PhysRevLett.48.359,PhysRevLett.62.2747}) within the unit cell \cite{RevModPhys.66.899,PhysRevB.96.245115}.
Next, we will use a similar method to show explicitly that the c.m. Zak phase is related to the c.m. position.
Firstly, the multi-particle Wannier function centered at $R_0$ for an isolated band can be defined as
\begin{equation}
\label{eq:wannier_function}
|{w_n}\left( R_0  \right)\rangle = \frac{1}{\sqrt{L}}\sum_K {{e^{-iKR_0}}|\psi_K^n\rangle},
\end{equation}
where $|\psi_K^n\rangle$ is the many-body Bloch state with respect to the c.m. momentum introduced in Eq.~\eqref{Def:ManyBody_BlochStates}.
Next, we introduce the c.m. position operator $ \hat R $ such that $\hat{R}| {r_1},{r_2},...,{r_N}\rangle= (\sum_i {{r_i}/N})|{r_1},{r_2},...,{r_N}\rangle$.
Then we would like to calculate the expectation value of $\hat R$ with respect to $|{w_n}\left( R_0 \right)\rangle$
\begin{eqnarray}
\label{wannier_expectation_R}
{\langle \hat{R}} \rangle_w
&&= \langle {w_n}\left( {{R_0}} \right)|\hat R|{w_n}\left( {{R_0}} \right)\rangle \nonumber \\
&&  = \frac{1}{L}\sum_{K,K'} e^{-i(K-K')R_0}\langle {\psi_{K'}^n}|\hat R|{\psi _K^n}\rangle.
\end{eqnarray}
To calculate $\langle {\psi_{K'}^n}|\hat R|{\psi _K^n}\rangle$, we adopt the argument in Ref.~\cite{PhysRevLett.80.1800}, where the average position $\langle x \rangle $ of the extended wave function under the PBC should be calculated as
\begin{equation}
\langle x\rangle  = \frac{L}{{2\pi }}{\rm{Im }}\left[ {\log {\rm{ }}{\langle \psi |{e^{i\delta K\hat x}}|\psi \rangle } } \right].
\end{equation}
in which $\delta K=2\pi/L$.
Therefore, the matrix element $\langle {\psi_{K'}^n}|\hat R|{\psi _K^n}\rangle $ should be modified as $\langle \psi _{K'}^n|\hat R|\psi _K^n\rangle  = \frac{L}{{2\pi }}{\mathop{\rm Im}\nolimits} \left[ {\log \left( {\langle \psi _{K'}^n|{e^{i\frac{{2\pi }}{L}\hat R}}|\psi _K^n\rangle } \right)} \right]$.
With some algebraic calculations, one has
\begin{eqnarray}
\label{derivation_Resta_Cal}
\langle \psi _{K'}^n|&&{e^{i\frac{{2\pi }}{L}\hat R}}|\psi _K^n\rangle = \frac{1}{L}\sum\limits_{l,l'} {{e^{i\left( {Kl - K'l'} \right)}}} \sum\limits_{\alpha,\alpha'} ({u_{K',\alpha'}^{n*}u^n_{K,\alpha} } \nonumber \\
 && \qquad \times {e^{i\left( {K{\Sigma _i}{r_i} - K'{\Sigma _i}{{r'}_i}} \right)/N}}\langle \{ r_{\alpha'}  \} + l'|{e^{i\delta K\hat R}}|\{ r_\alpha  \} + l\rangle) \nonumber \\
 &&= \frac{1}{L}\sum\limits_{l,l'} {{e^{i\left( {Kl - K'l'} \right)}}} \sum\limits_{{\alpha,\alpha'}} {{e^{i{\delta K(l+\Sigma_i{r_i/N})} }}} \nonumber \\
 && \qquad  \times {e^{i\left( {K{\Sigma _i}{r_i} - K'{\Sigma _i}{{r'}_i}} \right)/N}}  {u_{K',\alpha'}^{n*}u^n_{K,\alpha} } {\delta _{l,l'}}{\delta _{\alpha, \alpha '}} \nonumber \\
 &&= \frac{1}{L}\sum\limits_\alpha {u_{K',\alpha}^{n*}u^n_{K,\alpha} } \sum\limits_l  {{e^{i\left( {K - K' + \delta K} \right)(l  + {\Sigma _i}{{r}_i}/N)}}}\nonumber \\
 &&= \delta_{K',K+\delta K}\sum\limits_{{\alpha}} {u_{K',\alpha}^{n*}u^n_{K,\alpha} } ,
\end{eqnarray}
in which $\Sigma_i{r_i}$ is the summation of
Combining Eqs.~\eqref{wannier_expectation_R} and \eqref{derivation_Resta_Cal}, one obtains
\begin{eqnarray}
\label{eq:R_final1}
{\langle \hat R\rangle _w} &&= \sum\limits_K {\frac{1}{{2\pi }}{\mathop{\rm Im}\nolimits} \left[ {\log \left( {{e^{i\delta KR_0}}\sum\limits_{\alpha} {u_{K + \delta K,{\alpha}}^{n*}{u^n}_{K,{\alpha}}} } \right)} \right]} \nonumber \\
&& = {R_0} + \frac{1}{{2\pi }}{\rm{Im}}\left[ {\sum\limits_K {\log \langle u_{K + \delta K}^n|u_K^n\rangle } } \right] ,
\end{eqnarray}
in which we have used a compact notation $|u_K^n\rangle$, according to $\hat{H}(K)|u_K^n\rangle=E_K^n|u_K^n\rangle$.
Note that up to first order, there is $|u_{K + \delta K}^n\rangle  \approx |u_K^n\rangle  + {\partial _K}|u_K^n\rangle \delta K$.
Therefore, we have
\begin{equation}
\langle u_{K + \delta K}^n|u_K^n\rangle  \approx 1 + \langle {\partial _K}u_K^n|u_K^n\rangle \delta K.
\end{equation}
Using the relation $\log \left( {1 + x} \right) \approx x$ for $x\rightarrow0$, we can write  Eq.~\eqref{eq:R_final1} as
\begin{eqnarray}
\label{eq:R_final2}
{\langle \hat R\rangle _w}  &&\approx  {R_0} + \frac{1}{{2\pi }}{\rm{Im}}\left[ {\sum\limits_K {\log {\left(1 + \langle {\partial _K}u_K^n|u_K^n\rangle \delta K \right)} }} \right] \nonumber \\
&& \approx  {R_0} + \frac{1}{{2\pi }}{\rm{Im}}{\sum\limits_K {\langle {\partial _K}u_K^n|u_K^n\rangle \delta K}}   \nonumber \\
&& =  {R_0} + \frac{i}{{2\pi }}{\sum\limits_K {\langle u_K^n|{\partial _K}|u_K^n\rangle \delta K}} ,
\end{eqnarray}
in which we have used the fact that ${\langle {\partial _K}u_K^n|u_K^n\rangle}$ is a purely imaginary quantity so that ${\langle {\partial _K}u_K^n|u_K^n\rangle} = -\langle u_K^n|{\partial _K}|u_K^n\rangle$ .
In the thermodynamic limit $L\rightarrow \infty$, the summation Eq.~\eqref{eq:R_final2} becomes integral
\begin{eqnarray}
\label{eq:R_final3}
{\langle \hat R\rangle _w} &&= R_0 + \frac{i}{2\pi}\int_{ - \pi }^\pi  {\langle u_K^n|{\partial _K}|u_K^n\rangle dK}  \nonumber \\
&&= R_0 + \frac{\gamma _{{\rm{Zak}}}^n}{2\pi},
\end{eqnarray}
Therefore, we know that the c.m. Zak phase is related to the c.m. position of multi-particle Wannier states.
The result can be also generalized to 1D superlattice or model in higher dimension.
\subsection{Model}
\label{Sec:Model}
\begin{figure}[htp]
  \includegraphics[width = \columnwidth ]{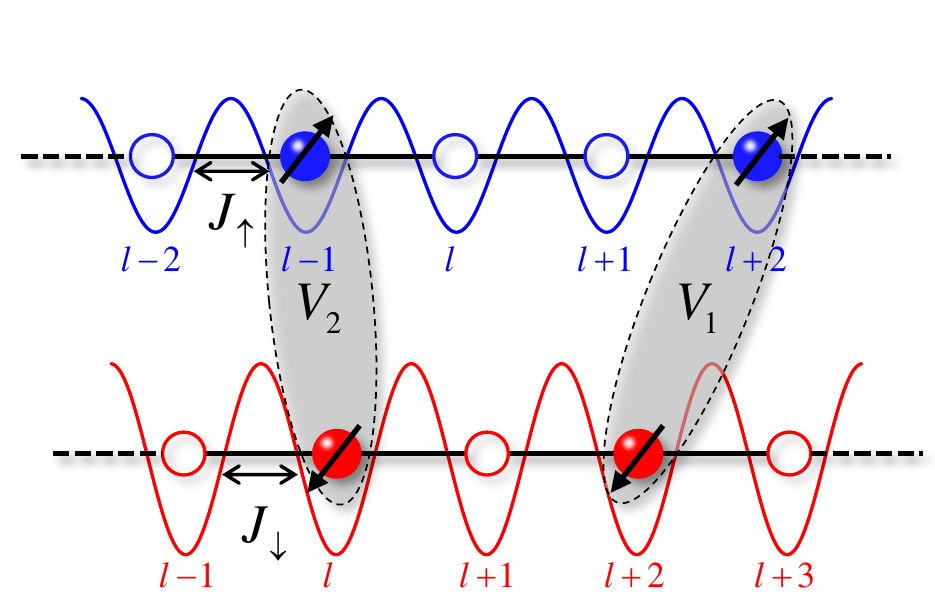}
  \caption{\label{fig:lattice}
  Sketch for 1D state-dependent lattice where two kinds of spin-1/2 particles are trapped separately.
  The phase of two periodic lattices have relative spatial shift.
  Blue and red lattices represent the periodic potential for spin-up and spin-down particles respectively.
  Arrows represent the NN tunneling.
  Grey shadings represent the NN inter-species interaction.
  The two lattices are overlapped in real space.
  The labels of sites are shown, and we will adapt this convention of labelling the sites in this work.
  }
\end{figure}

With the theory of describing the many-body topological properties in hand, we would like to explore the existence of interaction-induced topological states.
We consider two spin-1/2 particles trapped by a state-dependent optical lattices  \cite{PhysRevLett.82.1975,Olaf2003Controlled,PhysRevA.96.011602,PhysRevLett.91.010407}.
The state-dependent lattices considered here consists of two lattices, which have the same periodicity but different phase $\Delta \phi$.
These two different particles are trapped separately.
The phase difference between two lattices indicates they have a relative spatial shift in real space, as shown in Fig.~\ref{fig:lattice}.
The Hamiltonian of this system can be written as
\begin{eqnarray}
\label{eq:Ham}
\hat H &=& -\sum\limits_{l,\sigma  =  \uparrow , \downarrow } {\left( {{J_\sigma }\hat c_{\sigma ,l}^\dag {{\hat c}_{\sigma ,l + 1}} + H.c.} \right)}  \nonumber \\
&&+ \sum\limits_l {\left( {{{V_1\hat n}_{\downarrow,l}}{{\hat n}_{\uparrow,l}} + V_2{{\hat n}_{\downarrow,l + 1}}{{\hat n}_{\uparrow,l}}} \right)}
\end{eqnarray}
where $ \hat c_{\sigma , l}^\dag $ and $ \hat c_{\sigma , l}$  are respectively the creation and annihilation operators of spin-$\sigma$ particles in $l$-th site ($\sigma  =  \uparrow , \downarrow, \; l = 1,2, \cdots ,L$).
We suppose there are only the intra-species tunnelings $J_{\sigma}$, and the inter-species tunnelings (spin flip) are forbidden.
Thus, the particle number of each species is conserved.
Furthermore, we assume nearest-neighbor (NN) inter-species interaction between spin-up and spin-down particles are present.
The interaction strength will depend on the relative distance between the particles (see Appendix.~\ref{appendix:state_dependent_OL} for detailed discussions), and can be tuned via Feshbach resonance \citep{PhysRevLett.87.120406}.
As the two lattices have relative shift, there are two kinds of NN interaction strengthes $V_{1,2}$, as depicted by the grey-dashed lines in Fig.~\ref{fig:lattice}.
Without loss of generality, the interaction is assumed to be repulsive ($V_{1,2}>0$) by default in the following context.

\subsection{Solving the two-particle energy bands}
\label{Sec:Solution_BSs}
\begin{figure}
  \includegraphics[width = \columnwidth ]{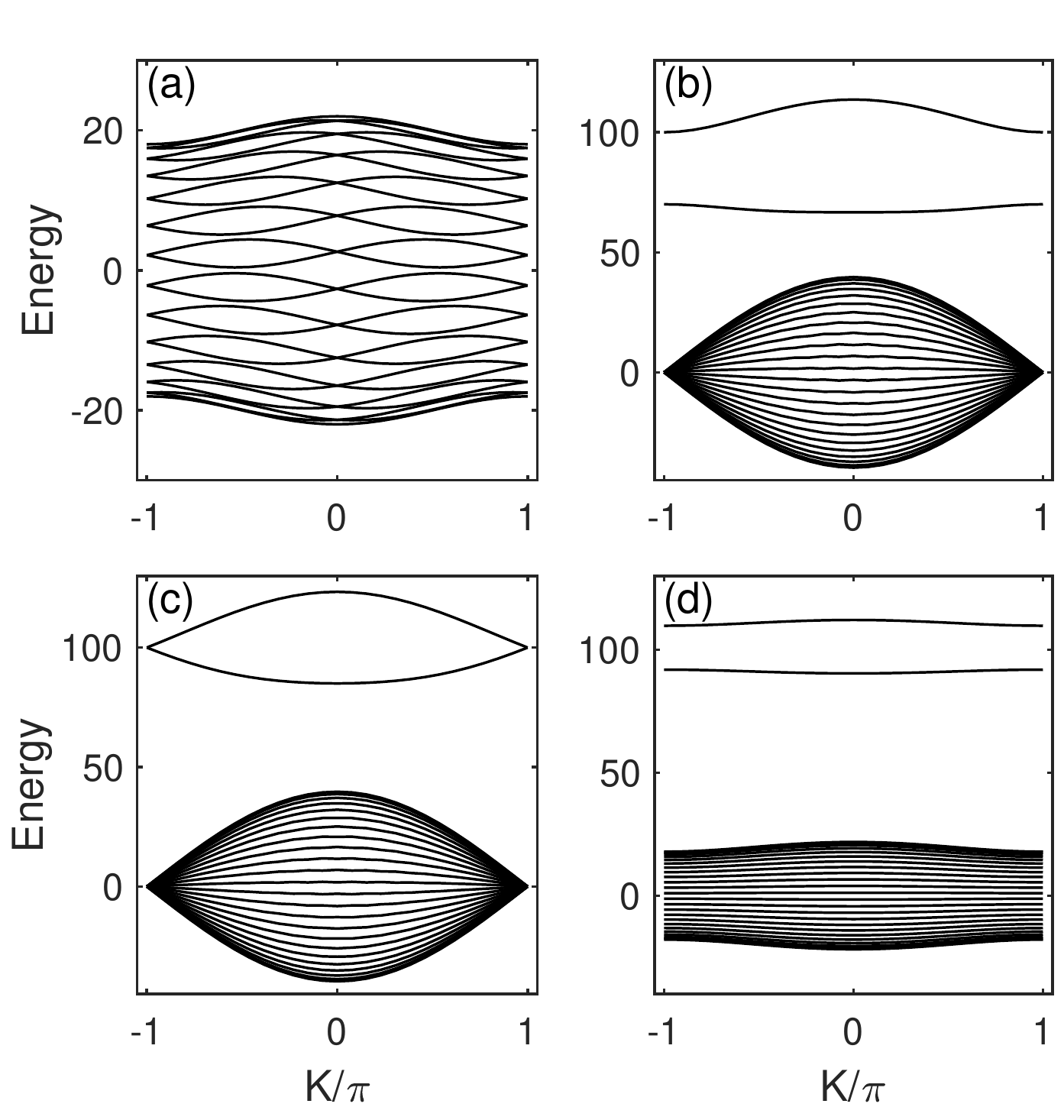}
  \caption{\label{fig:spectrum}
 Band structure with respect to c.m. quasi-momentum $K$ in four different conditions:
 (a) $J_\downarrow=1,J_\uparrow$ and $V=0$.
 (b) $J_\downarrow=J_\uparrow=$, but the two kinds of interaction differ: $V_1\neq V_2$.
 This condition corresponds to that the two sets of lattices have a relative shift $\Delta x \neq a/2$ so that the two interactions are imbalanced.
 (c) Gapless condition for isolated bands: $V_1 = V_2$ and $J_\downarrow = J_\uparrow$.
 (d) Topological condition: $V_1 = V_2$ and $J_\downarrow<J_\uparrow$.
 The $V_1=V_2$ condition corresponds to that the relative shift of two lattices is $\Delta x =a/2$.
 The size of system is set as $L_\downarrow=L_\uparrow=26$.
  }
\end{figure}

We first consider only two particles with different spin in this system, as this can be easily solved and may shed the light on some significant physics.
The two-particle subspace can be spanned by the following basis
\begin{equation}
{\cal H^{(\text 2)}} =\{ |{r_ \downarrow },{r_\uparrow}\rangle \},
\end{equation}
where $r_\downarrow$ ($r_\uparrow$) refers to the position of a (b) particle.
Considering lattice with $L_\downarrow=L_\uparrow=L$ sites, the dimension of the Hilbert space is $L^2$.
According to Sec.~\ref{Sec:General_Theory}, the eigenstates of cotranslation operator under periodic boundary condition are found to be
\begin{equation}
\label{eq:co_trans_eignstates}
|K,{r_{\uparrow\downarrow} }\rangle  = \frac{1}{{\sqrt L }}\sum\limits_{l = 1}^N {{e^{iKl}}|{r_ \downarrow } + l,{r_\uparrow } + l\rangle }
\end{equation}
where ${r_\downarrow } = {r_ \uparrow } + {r_{ \uparrow\downarrow}}$.
To be explicit, we have used the notation $r_{ \uparrow\downarrow}$ to label the eigenstates of cotranslation operator according to relative distance between two particles.
This is a convenient way to distinguish the seed basis.
As discussed in Sec.~\ref{Sec:General_Theory}, the choice of $r_\downarrow$ and $r_\uparrow$ determines the gauge.
In this calculation, we fix $r_\uparrow=0$ and let $r_\downarrow=r_{\downarrow\uparrow}$.
Then the seed basis used here are $\left\{ {|{r_{\downarrow\uparrow}},0\rangle } \right\},\;{r_{\downarrow\uparrow}} \in \left[ { - L/2,L/2} \right]$.
The Bloch Hamiltonian can be derived from (\ref{eq:Ham}) as
\begin{equation}
{H_{r'_{\downarrow\uparrow},r_{\downarrow\uparrow}}}(K) = \langle K,r'_{\downarrow\uparrow}|\hat H|K,r_{\downarrow\uparrow}\rangle.
\end{equation}
which is a $L$-by-$L$ matrix in the basis of $|K,r_{\downarrow\uparrow}\rangle$.
For convenience, we arrange the basis in the order of $\{|K,r_{\downarrow\uparrow}\rangle\}$ by $r_{\downarrow\uparrow} = 1, 2\ldots,N/2 - 1, - N/2, - N/2 + 1,\ldots, - 1, 0$.
In this manner, the matrix representation of $H(K)$ reads
\begin{widetext}
\begin{equation}
\label{eq:Ham_momentum_matrix}
H\left( K \right) = \left[ {\begin{array}{*{20}{c}}
V_2&{ - \left( {{J_\downarrow} + {J_\uparrow}{e^{ - iK}}} \right)}&0& \cdots &{ - \left( {{J_\downarrow} + {J_\uparrow}{e^{iK}}} \right)}\\
{ - \left( {{J_\downarrow} + {J_\uparrow}{e^{iK}}} \right)}&0&{ - \left( {{J_\downarrow} + {J_\uparrow}{e^{ - iK}}} \right)}& \cdots &0\\
0&{ - \left( {{J_\downarrow} + {J_\uparrow}{e^{iK}}} \right)}& \ddots & \cdots & \vdots \\
 \vdots & \vdots & \vdots &0&{ - \left( {{J_\downarrow} + {J_\uparrow}{e^{ - iK}}} \right)}\\
{ - \left( {{J_\downarrow} + {J_\uparrow}{e^{ - iK}}} \right)}&0&0&{ - \left( {{J_\downarrow} + {J_\uparrow}{e^{iK}}} \right)}&V_1
\end{array}} \right]
\end{equation}
\end{widetext}
By numerically diagonalizing the Bloch Hamiltonian Eq.~\eqref{eq:Ham_momentum_matrix} for all c.m. quasi momentum $K$, we obtain the energy bands of the system within first Brillouin zone.
In the non-interacting case, the system is simply the direct product of two normal lattice.
Correspondingly, the c.m. quasi momentum is an average of two single-particle momentum $K=(k_\downarrow+k_\uparrow)/2$.
The energy bands with respect to $K=(k_\downarrow+k_\uparrow)/2$ is shown in Fig.~\ref{fig:spectrum} (a).
For strong interaction, there appear one continnum and two isolated bands, as shown in Fig.~\ref{fig:spectrum} (b-d).
The continuum corresponds to states that two particles move quasi independently.
The isolated bands correspond to bound states, where particles are bound by interaction and perform correlated dynamics.
The appearance of two isolated bands is because there are two kinds of interactions $V_{1,2}$.
We find that when $V_1\neq V_2$, or $V_1=V_2$ and $J_\downarrow\neq J_\uparrow$, there is a gap between the isolated bands, as shown in Fig.~\ref{fig:spectrum} (b) and (d), respectively.
However, when $J_\downarrow=J_\uparrow$ and $V_1=V_2$, the two isolated bands become gapless, see Fig.~\ref{fig:spectrum} (c).

\subsection{Topological nature of isolated bands}
\label{Sec:Topological_BSs}
As pointed in the previous subsection, the system is the direct product of two simple lattice under interaction-free condition.
It is already known that this system is completely topologically trivial.
For strongly interacting condition, the two isolated bands are away from the continuum band.
Since the two particles within the continuum band are quasi independent, and the lattices are both trivial, the continuum states should also be trivial.
It is more intriguing to explore the topological properties of isolated bands.

\subsubsection{Discrete symmetry and quantized Zak phase}
\begin{figure}
  \includegraphics[width = \columnwidth ]{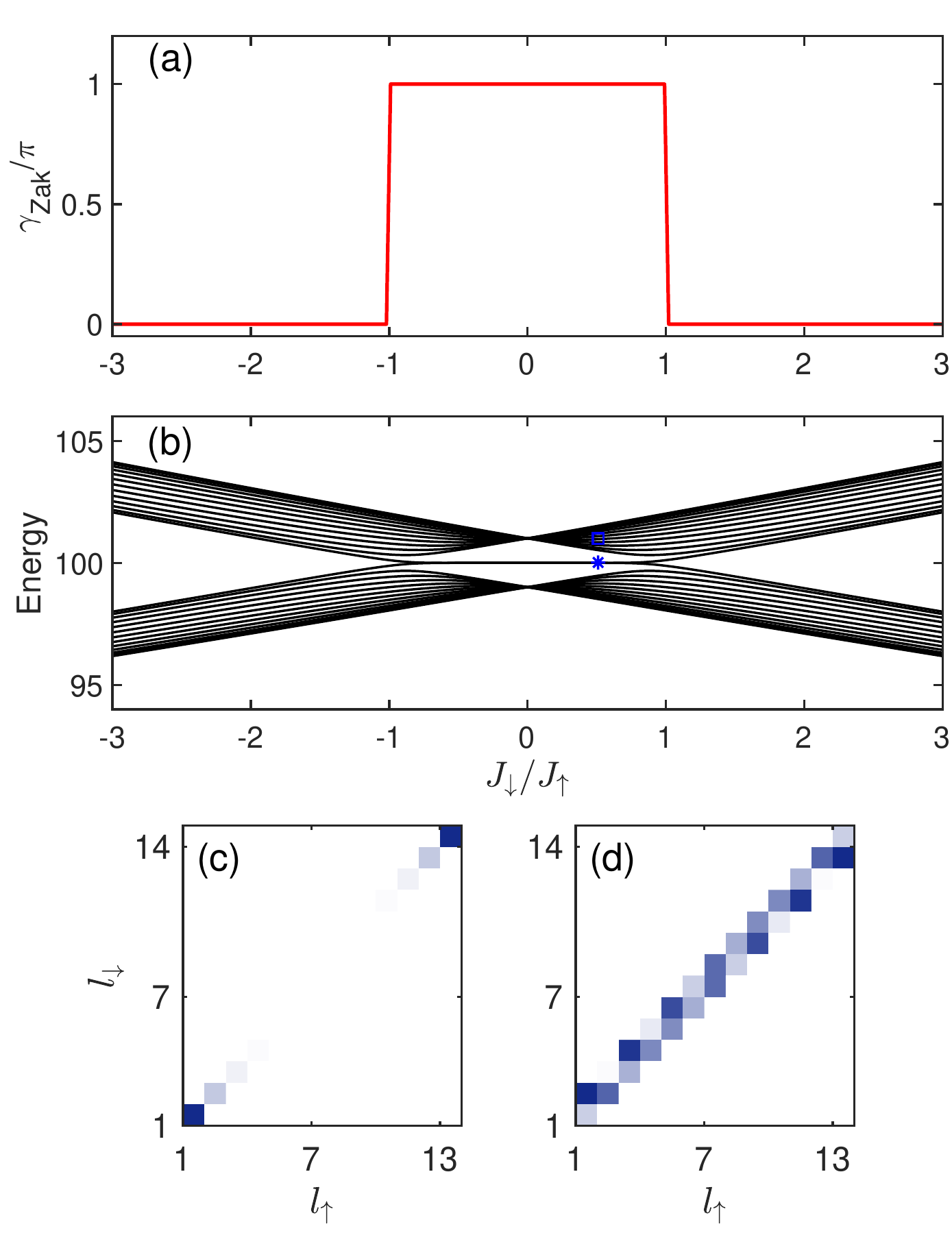}
  \caption{\label{fig:EdgeBSs}
 (a) The Zak phase of upper isolated band calculated versus $J_\downarrow/J_\uparrow$ (with fixed $J_\uparrow=1$ in calculation).
 (b) Eigenenergies of bound states versus $J_\downarrow/J_\uparrow$.
 The energy spectrum of scattering states are not shown here.
 Blue star and square respectively mark the in-gap bound states and the bulk bound states at $J_\downarrow/J_\uparrow = 1/2$, and their correlation matrices $\Gamma_{{l,l'}}  = \langle {n_{\downarrow,l}}{n_{{\uparrow,l' }}}\rangle$ are shown in (c) and (d).
 The length of system is $L_\downarrow=L_\uparrow=14$ for (a) with PBC, and $L_\downarrow=14$, $L_\uparrow=13$ for (b-d) with symmetric OBC.
 The strength of interaction is set to $V=100$.
 }
\end{figure}

We note that, if $V_1=V_2=V$, the Bloch Hamiltonian Eq.~\eqref{eq:Ham_momentum_matrix} possesses the \emph{c.m. inversion symmetry}
\begin{equation}
\label{eqn:cm_inversion_sym}
{\mathcal{I}}H(K){\mathcal{I}}^{-1}=H(-K),
\end{equation}
where
\begin{equation}
{\cal I} = \left[ {\begin{array}{*{20}{c}}
{}&{}&{}&1\\
{}&{}& {\mathinner{\mkern2mu\raise1pt\hbox{.}\mkern2mu
 \raise4pt\hbox{.}\mkern2mu\raise7pt\hbox{.}\mkern1mu}} &{}\\
{}& {\mathinner{\mkern2mu\raise1pt\hbox{.}\mkern2mu
 \raise4pt\hbox{.}\mkern2mu\raise7pt\hbox{.}\mkern1mu}} &{}&{}\\
1&{}&{}&{}
\end{array}} \right],
\end{equation}
and $\mathcal{I} ^2=\bold{1}$.
The c.m. inversion symmetry manifests that the system is invariant if the relative distance of two particles $ r_{\downarrow\uparrow} = r_\downarrow-r_\uparrow$ are reversed $r_{\downarrow\uparrow} \rightarrow -r_{\downarrow\uparrow}$.
This is the results of symmetry of interaction and the cotranslation symmetry.
The eigenstates of $H(K)$ and $H(-K)$ are therefore connected via
\begin{equation}
{\cal I}|u_K^n\rangle {\rm{ = }}{{\rm{e}}^{i\theta \left( K \right)}}|u_{ - K}^n\rangle,
\end{equation}
where ${\theta(K)}$ is a $K$-dependent function with $\theta (K+2\pi)=\theta (K) + 2\pi m$.
Therefore, one can obtain the following relation
\begin{eqnarray}
A_{ - K}^n &&= \langle u_{ - K}^n|{\partial _{ - K}}|u_{ - K}^n\rangle \nonumber \\
 &&=  - \langle u_K^n|{e^{  i\theta (K)}}{{\cal I}^{ - 1}}{\partial _K}\left( {{e^{-i\theta (K)}}{\cal I}|u_K^n\rangle } \right) \nonumber \\
 &&=   i{\partial _K}\theta \left( K \right) - \langle u_K^n|{\partial _K}|u_K^n\rangle \nonumber \\
 &&=   i{\partial _K}\theta \left( K \right) - A_K^n,
\end{eqnarray}
and hence the Zak phase is given as
\begin{eqnarray}
\gamma _{{\rm{Zak}}}^n &&=   i\int\limits_{ - \pi }^\pi  {A_K^ndK}  =   i\int\limits_{ - \pi }^\pi  {\left( {  i{\partial _K}\theta \left( K \right) - A_{ - K}^n} \right)dK}  \nonumber \\
&&= - [\theta \left( \pi  \right) - \theta \left( {  -\pi } \right)] - i\int\limits_{ - \pi }^\pi  {A_{ - K}^nd { K}}  \nonumber \\
&&= 2\pi m - \gamma _{{\rm{Zak}}}^n, \quad m \in \mathbb{Z},
\end{eqnarray}
which implies that  the Zak phase of the system is quantized to $0$ or $\pi$ mod $2\pi$ \cite{PhysRevLett.62.2747,PhysRevB.88.245126}.
It is quite easy to understand why the Zak phase is quantized by considering the relation between Zak phase and the c.m. position of multi-particle Wannier states.
The c.m. inversion symmetry requires that the c.m. position of the multi-particle Wannier states should be centered at either of the two inversion-symmetric points of the lattice.
This can be seen from Eq.~\ref{eq:R_final2} that, since the Zak phase is quantized to $0$ or $\pi$ mod $2\pi$, the center of multi-particle Wannier state ${\langle \hat R\rangle _w} = R_0 + {\gamma _{{\rm{Zak}}}^n}/{2\pi}$ would take half-integer value, indicating it is only centered at inversion-symmetric points.
The discussion on imbalanced interaction $V_1\neq V_2$ is presented in Appendix.~\ref{appendix:imbalance}.
In addition, the system also possesses the time-reversal symmetry
\begin{equation}
{\mathcal K}H\left( K \right){{\mathcal K}^{ - 1}} = H\left( { - K} \right),
\end{equation}
in which $\mathcal{K}$ is the complex-conjugation operator.
As shown in the previous section, the isolated bands of this system are well separated from continuum as long as the interaction is strong enough, and they are still gapped from each other in the limit of thermodynamic if $J_\downarrow \neq J_\uparrow$.
We can therefore evaluate the Zak phase of the isolated bands according to Eq.~\eqref{Def:ZakPhase}.
We present the numerical calculation of the Zak phase according to Eq.~\eqref{Def:ZakPhase} versus different ratios of $J_\downarrow/J_\uparrow$ in Fig.~\ref{fig:EdgeBSs} (a).
The result shows a great quantization of Zak phase, and the phase transition at $|J_\downarrow/J_\uparrow| = 1$ is very distinct.
Therefore, we are able to identify two topologically different phases.
If $|J_\downarrow/J_\uparrow|>1$, the system is trivial, with the Zak phase $\gamma_{\rm{Zak}}=0$.
If $|J_\downarrow/J_\uparrow|<1$, the system is topological, with the Zak phase $\gamma_{\rm{Zak}}=\pi$.
The gapless condition $|J_\downarrow|=|J_\uparrow|$ is characterized as the topological transition point.
For $V_1\neq V_2$,
It is known that, in the conventional topological band theory, the Zak phase is affected by the choice of unit cell \cite{PhysRevB.95.035421} and the gauge of Fourier transformation \citep{PhysRevB.91.125424}.
As stated in Sec.~\ref{Sec:General_Theory}, the choice of seed basis affects the c.m. Zak phase.
For example, if instead one chooses a different kind of seed basis $r_\downarrow \equiv 0$, $r_\uparrow\equiv r_{\downarrow\uparrow}  = 0, 1,\ldots,N/2 - 1, - N/2, - N/2 + 1,\ldots, - 1$, the off-diagonal matrix elements $-(J_\downarrow+J_\uparrow e^{-iK})$ in Eq.~\eqref{eq:Ham_momentum_matrix} would be changed to $-(J_\uparrow+J_\downarrow e^{-iK})$.
In this gauge, there is $\gamma_{\rm{Zak}}=0$ if $|J_\downarrow/J_\uparrow|<1$ and $\gamma_{\rm{Zak}}=\pi$ if $|J_\downarrow/J_\uparrow|>1$.
However, the difference between these two phases is $\delta \gamma_{\rm{Zak}} = \pi$ mod $2\pi$, which is independent of the choice of gauge.
In other words, these two phases are still topologically distinct in this gauge choice.
In the next section Sec.~\ref{Sec:Pumping}, we will present the topological pumping, which will further prove there are two distinct phases.
Effective single-particle model for the isolated bands is derived up to second order in Appendix.~\ref{appendix:eff_model}.
The effective model shows a zigzag geometry, which still preserves the similar inversion symmetry.

\subsubsection{Bulk-edge correspondence}
Keeping $V_1=V_2=V$, we proceed to investigate the existence of topological edge bound-states by imposing the open boundary condition (OBC).
In fact, there are two different strategies to determine how the edge is terminated.
One kind of the OBC is that $L_\downarrow=L_\uparrow$, where the edge breaks the c.m. inversion symmetry, and we shall mention it by the \emph{asymmetric boundary}.
Another kind of the OBC, mentioned by symmetric boundary, is that $L_\downarrow=L_\uparrow+1$, which preserves the c.m. inversion symmetry.
In the main text, we only consider the symmetric boundary.
For more discussions about these two kinds of terminations, see Appendix.~\ref{appendix:OBC}.
By using exact diagonalization, we calculate the energy spectrum for different ratios of $J_\downarrow/J_\uparrow$ under the symmetric open boundary condition.
We find that there are doubly degenerate in-gap states between the bulk of bound states only for $|J_\downarrow/J_\uparrow|<1$, as shown in Fig.~\ref{fig:EdgeBSs} (b).
Compared with Fig.~\ref{fig:EdgeBSs} (a), it is clear that this is in great agreement with the bulk topology discussed above.
To uncover the properties of the in-gap bound states, we calculate the correlations of the two particles $\Gamma_{{l,l'}}  = \langle {n_{\downarrow,l}}{n_{{\uparrow,l' }}}\rangle$.
The correlation pattern shown in Fig.~\ref{fig:EdgeBSs} (c) clearly reveals the in-gap states are strongly localized and correlated, indicating the existence of edge bound-states.
For comparison, the bound states in bulk of band are highly delocalized but still bound together, see Fig.~\ref{fig:EdgeBSs} (d).
The appearance and disappearance of in-gap bound states at edges is in agreement with our analysis in the bulk topology.
As stated in Sec.~\ref{Sec:General_Theory}, the c.m. Zak phase is related to the c.m. position of multi-particle Wannier states within a cell.
Non-trivial Zak phase leads to extra density accumulation at the terminated edge if the edge is commensurate to the c.m. inversion symmetry.
This phenomenon is usually called the bulk-edge correspondence \cite{RevModPhys.82.3045,PhysRevB.88.245126}.

\begin{figure*}[htp]
  \includegraphics[width = \textwidth ]{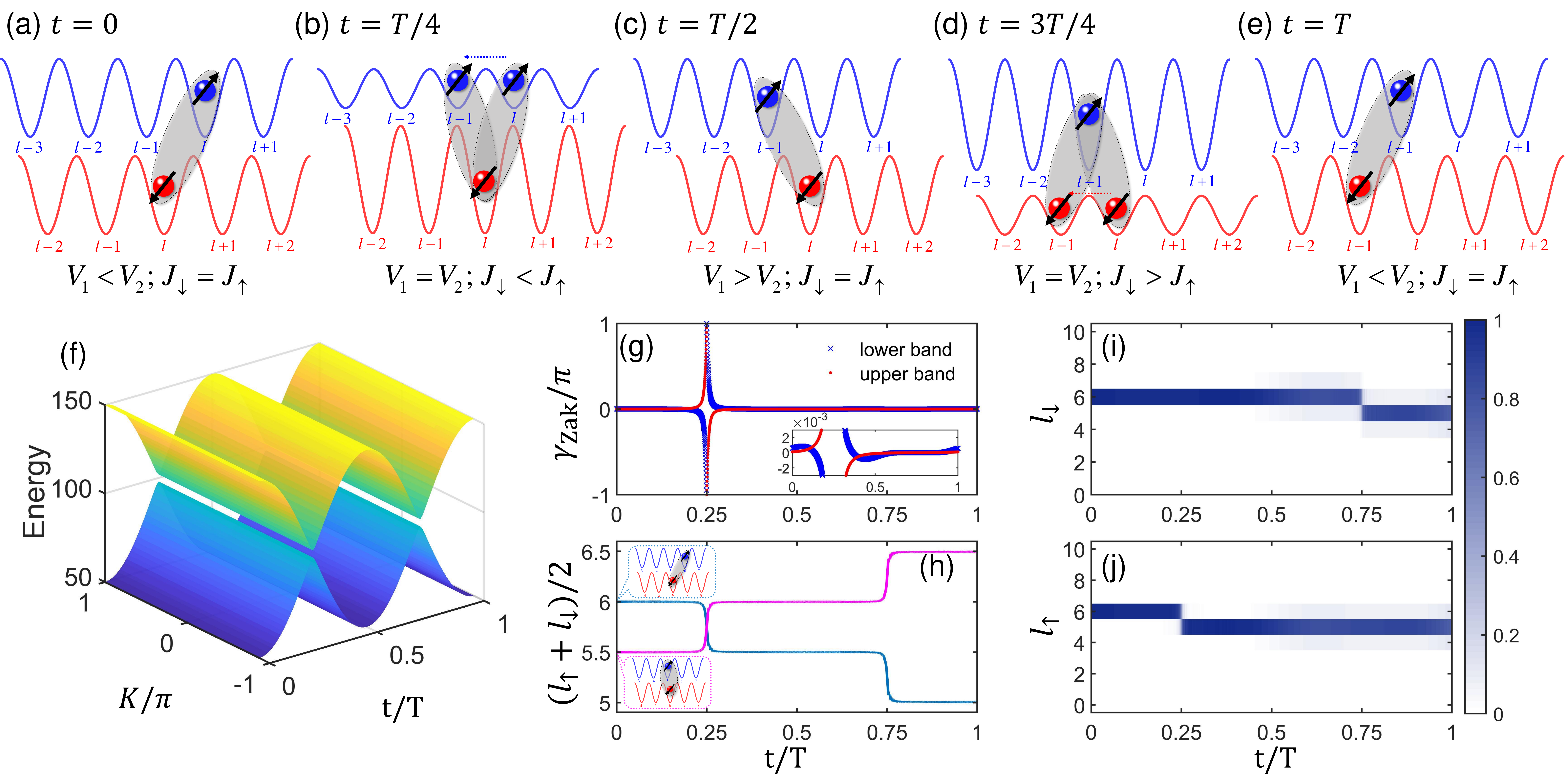}
  \caption{\label{fig:Pumping_Scheme}
 Topological pumping of two-particle bound state in one period $T$.
 (a-e) Schematic illustrations of the pumping process.
 Red and blue lattices refer to periodic potentials for particle $\downarrow$ and $\uparrow$ respectively.
 The bound state is formed as long as the interaction is strong enough.
 In (b) and (d), there appears degenerate states due to the equal strength of interaction $V_1=V_2$ .
 (f) A 3D view of bound-state energy bands in closed $K-t$ space of two-particle system.
 (g) The Zak phase $\gamma_\mathrm{Zak}$ of lower and upper isolated bands as a function of time $t$ in a period.
 Inset shows an enlarged area around $\gamma_\mathrm{Zak}=0$.
 (h-j) Numerical simulation of two-particle topological pumping.
 (h) shows a c.m. position shift of the two particles $l_{\rm{c.m.}} = (l_\downarrow +l_\uparrow)/2$ starting from two different initial states: $|\psi_{i_1}(0)\rangle = |6_\downarrow, 6_\uparrow \rangle$ (blue line) and $|\psi_{i_2}(0)\rangle = |6_\downarrow, 5_\uparrow \rangle$ (pink line).
 Insets show the schematic diagram of the two initial states.
 (i-j) show density distributions of the two particles respectively during the pumping.
 For simplicity, we only show one of the results initialized as $|\psi_{i_1}\rangle = |6_\downarrow, 6_\uparrow \rangle$.
 Pumping parameters are chosen as $J=1, V = 100, \Delta_0=50, \delta_0=1$ and $\omega = 3\times 10^{-4}, \phi_0=0$.
 Lattice length is $L_\downarrow=L_\uparrow=11$.
 }
\end{figure*}

\section{Interaction-induced Thouless pumping}
\label{Sec:Pumping}
In this section, we would like to explore the topological pumping of bound states induced by the interaction effect.
Recalling that by adding modulation of on-site energy and tunneling, the SSH model is extended to the Rice-Mele model \cite{PhysRevLett.49.1455}.
In the topological Thouless pumping scheme \citep{PhysRevB.27.6083, PhysRevLett.64.1812}, the two distinct topological phases are connected through breaking the chiral (inversion) symmetry without closing the energy gap.
After one period of pumping, the Zak phase winds for $2\pi$, and the polarization is changed for a quanta.
The Thouless pumping has already been realized experimentally and quantized particle/charge transport is observed \cite{lohse2016thouless, nakajima2016topological}.
For interacting system, the topological pumping of two interacting bosons \cite{PhysRevA.95.063630} and many-body case \cite{PhysRevLett.106.110405, PhysRevB.98.245148} has been investigated.
However, these interacting models have the topological single-particle counterpart in the interaction-free condition.
Here, we propose a scheme to realize the quantized particle transport based on the interaction-induced topological bound states discussed in previous section.
In the same spirit of topological Thouless pumping, we add the modulations of both interaction and tunneling terms into our model, which are given as
\begin{eqnarray}
{{\hat H}_J}\left( t \right) &&=  - \left( {J - \delta \left( t \right)} \right)\sum\limits_l {\left( {\hat c_{\downarrow, l}^\dag {{\hat c_{\downarrow, l+1}}} + H.c.} \right)} \nonumber \\
&&- \left( {J + \delta \left( t \right)} \right)\sum\limits_l {\left( {\hat c_{\uparrow, l}^\dag {{\hat c}_{\uparrow, l + 1}} + H.c.} \right)},
\end{eqnarray}
with $\delta(t)=\delta_0\sin (\omega t+\varphi_0)$, and
\begin{eqnarray}
{{\hat H}_V}\left( t \right) &&= \left( {V - \Delta \left( t \right)} \right)\sum\limits_l {{{\hat n}_{\downarrow,l}}{{\hat n}_{\uparrow,l}}}  \nonumber \\
&&+ \left( {V + \Delta \left( t \right)} \right)\sum\limits_l {{{\hat n}_{\downarrow,l + 1}}{{\hat n}_{\uparrow,l}}} ,
\end{eqnarray}
with $\Delta(t) = \Delta_0 \cos (\omega t +\varphi_0)$.
Here, $\delta_0$ and $\Delta_0$ are the modulation strengthes of hopping and interactions,  $\omega$ is the common modulation frequency, and $\phi_0$ is the initial phase.
The full modulated Hamiltonian is $\hat H (t)={{\hat H}_J} (t)+ {{\hat H}_V} (t)$.
Experimentally, the modulation of tunneling can be realized through modulating the height of periodic potential.
For the modulation of interaction, it can be implemented by tuning relative position of the two periodic potential \cite{PhysRevLett.91.010407}, where the interaction is assumed to depend on the relative distance between particles.
The demonstration of the pumping process is shown in Fig.~\ref{fig:Pumping_Scheme} (a-e).
Note that our pumping scheme is essentially different from the coherent transport using state-dependent lattice in Ref.~\cite{PhysRevLett.91.010407}.
Under the PBC, our pumping scheme adiabatically connects two distinct topological phases of bound states via breaking the c.m. inversion symmetry without closing the gap.
As shown in Sec.~\ref{Sec:General_Theory}, the c.m. Zak phase indicates the c.m. position within a unit cell.
After a pumping cycle, the Zak phase changes $2\pi$, and correspondingly, the c.m. position of particles are shifted for a unit cell.
This is also a useful approach to justify the topological nature.
Consider $\varphi_0=0, T = 2\pi/\omega$, the resonant tunneling between the two nearest-neighboring positions mainly occurs at $t=T/4$ and $t=3T/4$ during the pumping cycle, which is protected by the non-trivial topology \cite{PhysRevB.84.195410}.
%

To ensure the existence of energy gap between isolated bands and continuum band, we focus on the regime that $|V \pm \Delta_0|\gg |J\pm \delta_0|$.
The bound-state energy bands in a closed $K-t$ space are shown in Fig.~\ref{fig:Pumping_Scheme} (f).
According to Eq.~\eqref{Def:ChernNum}, we calculate the Chern number of these two bands numerically, and the results are $C= \pm 1$ for upper and lower bands, respectively.
We also calculate the Zak phase as a function of time in Fig.~\ref{fig:Pumping_Scheme} (g).
With our choices of seed basis and the pumping parameters, the Zak phase will reach to $\pi$ at $t=T/4$ and to $0$ at $t=3T/4$.
At these two points, the system reduces to the model Eq.~\eqref{eq:Ham}, and the Zak phase is strictly quantized.
For other cases, the Zak phase is not quantized and will change with the parameters.
The winding of Zak phase for each band matches the associated Chern number, coinciding with the following formula \citep{RevModPhys.66.899}
\begin{equation}
\label{eqn:Chern_ZakPhase}
{C_n} = \frac{1}{2\pi}\int\limits_0^T {dt\frac{\partial }{{\partial t}}\gamma _{{\rm{Zak}}}^n\left( t \right)} .
\end{equation}
Since the Zak phase is affected by the choice of seed basis and origin of unit cell \citep{PhysRevB.95.035421, PhysRevX.8.021065}, the results in Fig.~\ref{fig:Pumping_Scheme} (g) is not unique, and may vary with different choices \citep{PhysRevX.8.021065}.
However, the winding of Zak phase in Eq.~\eqref{eqn:Chern_ZakPhase} after a full pumping cycle is invariant and quantized \citep{RevModPhys.66.899, PhysRevB.47.1651, PhysRevB.96.245115}.
As a verification of the topological pumping, we further simulate the quasi-adiabatic pumping numerically.
We shall start with two interacting particles in nearest-neighboring site as an initial bound state:  $|\psi_{i_1}(0)\rangle = |6_\downarrow, 6_\uparrow \rangle$ or $|\psi_{i_2}(0)\rangle = |6_\downarrow, 5_\uparrow \rangle$.
These two kinds of initial bound states are dominated respectively by the two kinds of interaction $V_1$ and $V_2$.
During the adiabatic pumping, these two kinds of initial states will evolve along lower or upper isolated bands accordingly.
Thus, the results of c.m. position shift will reflect the topology of the bands.
The c.m. position of the particles during the time evolution are presented in Fig.~\ref{fig:Pumping_Scheme}  (h).
We also present the density distributions of particles initialized in $|\psi_{i_1}(0)\rangle = |6_\downarrow, 6_\uparrow \rangle$ during the pumping in Fig.~\ref{fig:Pumping_Scheme}  (i-j).
The results of numerical simulation show that the c.m. position of particles is shifted for one unit towards left after a complete period.
The change of c.m. position $(\Delta l_\downarrow + \Delta l_\uparrow)/2 =+1/-1$ is in agreement with the Chern number of upper/lower band.

\section{Summary and discussions}
\label{Sec:Summary}

In summary, we have investigated the topological nature of interacting bound states and their transport in a state-dependent lattice.
We find the existence of topological bound states protected by the c.m. inversion symmetry.
This kind of symmetry requires the system should be invariant under the interchange of two kinds of distinguishable particles.
The topological nature of bound states can be well characterized by the quantized c.m. Zak phase.
As a result of bulk-edge correspondence, there appear topological edge bound-states corresponding to nontrivial c.m. Zak phases.
The repulsive bound pair can be regarded as the two-hole excitation of a filled attractively interacting system.
Therefore, to some extent, the topological bound states reflect the topological excitations in interacting many-body quantum system.
Furthermore, the c.m. inversion symmetry in our system may suggest the possible existence of symmetry-protected (SPT) phase \citep{RevModPhys.89.041004} for the many-body ground state.

It should be noted that we assume no coherent population transfer between two internal states.
This is essential for realizing the c.m. inversion symmetry.
If there is such kind of population transfer, two particles become indistinguishable.
Therefore, the c.m. inversion symmetry does no longer exist, since exchanging the relative position of two indistinguishable particles has no physical meaning.
Without this essential symmetry, the Zak phase will no longer stay quantized.
This means the coherent population transfer is a symmetry-breaking term.

On the other hand, we have also proposed a topological Thouless pumping via periodically modulating interaction and tunneling simultaneously.
We obtain a non-trivial c.m. Chern number which is evidenced by the quantized shift of the c.m. position in the pumping process.
Moreover, although both systems involve state-dependent lattices, our topological transport is different from the coherent transport via shifting the potential~\citep{PhysRevLett.91.010407}.
In our scheme, the periodic potential is assumed to shift back and forth, instead of  shifting monotonously.
The interplay between interaction and tunneling plays important role during the pumping.
In future, it is intriguing to investigate the pumping of ground state via density matrix renormalization group or other techniques.
Our model may be realized in current cold atoms experiment.
The state-dependent optical lattice is an ideal platform \citep{PhysRevLett.91.010407, PhysRevLett.82.1975} in which the interaction and tunneling are highly controllable.
Besides, there are some other systems or techniques being the possible candidates for realizing the interaction-induced topological bound states, such as the synthetic zigzag optical lattice \citep{PhysRevA.94.063632, PhysRevA.87.033609}.
We also note that our model is similar to that in Ref.~\citep{PhysRevLett.116.225303}, and therefore the experimental consideration may be shared.

\acknowledgements{This work is supported by the Key-Area Research and Development Program of GuangDong Province under Grants No. 2019B030330001, the National Natural Science Foundation of China (NNSFC) under Grants No. 11874434 and No. 11574405, and the Science and Technology Program of Guangzhou (China) under Grants No. 201904020024.
Y.K. was partially supported by the Office of China Postdoctoral Council under Grant No. 20180052 and  the National Natural Science Foundation of China (NNSFC) under Grant No.11904419. }

\appendix
\section{State-dependent optical lattice}
\label{appendix:state_dependent_OL}
State-dependent lattice has been widely used in cold atoms experiment.
The intra-species interaction is used to control and produce the desired entanglement in quantum computation\citep{mandel2003controlled,PhysRevLett.91.107902}.
Usually, the state-dependent lattice consists of two $\sigma^+$ and $\sigma^-$ polarized standing waves.
Atoms with different internal states will be trapped by different standing waves.
The Hamiltonian of 1D state-dependent lattice with two bosonic components can be written in language of second-quantization
\begin{eqnarray}
\hat{H} &=&  \sum\limits_{\sigma  =  \uparrow , \downarrow } {\int d x\hat \Psi _\sigma ^\dag \left( x \right)} \left[ { - \frac{{{\hbar ^2}}}{{2m}}{\nabla ^2} + \mathcal{V}_\sigma ^{{\rm{lat}}}\left( x,\phi \right)} \right]{{\hat \Psi }_\sigma }\left( x \right)\nonumber \\
&+&{g_{{\rm{inter}}}}\int dx\hat{\Psi}_ \uparrow ^{{\rm{\dag}}}\left( {x} \right)\hat{\Psi}_ \downarrow ^{{\rm{\dag}}}\left( {x} \right){\hat{\Psi} _ \uparrow }\left( {x} \right){\hat{\Psi}_ \downarrow }\left( {x} \right)
\end{eqnarray}
where $\hat{\Psi} _\sigma $ is the field operator of spin-$\sigma\; (\sigma=\uparrow , \downarrow)$ particle,  $\mathcal{V}_\sigma ^{\rm{lat}}\left( x,\phi \right)$ is the state-dependent optical lattice potential,  and $g_\mathrm{inter}=4\pi a_s \hbar^2/m$ is the inter-species coupling constant determined by $s$-wave scattering process.
We have set lattice constant to be $a=1$.
As we only consider the inter-species interaction in our model, the intra-species interaction terms have been omitted.
The state-dependent optical lattice potential can be written as
\begin{eqnarray}
\label{eqn:spin_dependent_optical_potential}
\mathcal{V}_\uparrow ^{\rm{lat}} \left( x,\phi \right)=\mathcal{V}_\uparrow^{(0)}\cos^2(k_L x + \phi/2) \nonumber \\
\mathcal{V}_\downarrow ^{\rm{lat}} \left( x,\phi \right)= \mathcal{V}_\downarrow^{(0)}\cos^2(k_L x - \phi/2)
\end{eqnarray}
where $\mathcal{V}_\sigma^{(0)}$ is the height of lattice potential, $\phi$ is the relative phase shift of lattice, which is affected by the laser polarization and can be controlled via electro-optical modulator (EOM) experimentally \citep{PhysRevLett.91.010407,PhysRevA.96.011602,Belmechri_2013}.
\begin{figure}[htp]
  \includegraphics[width = \columnwidth ]{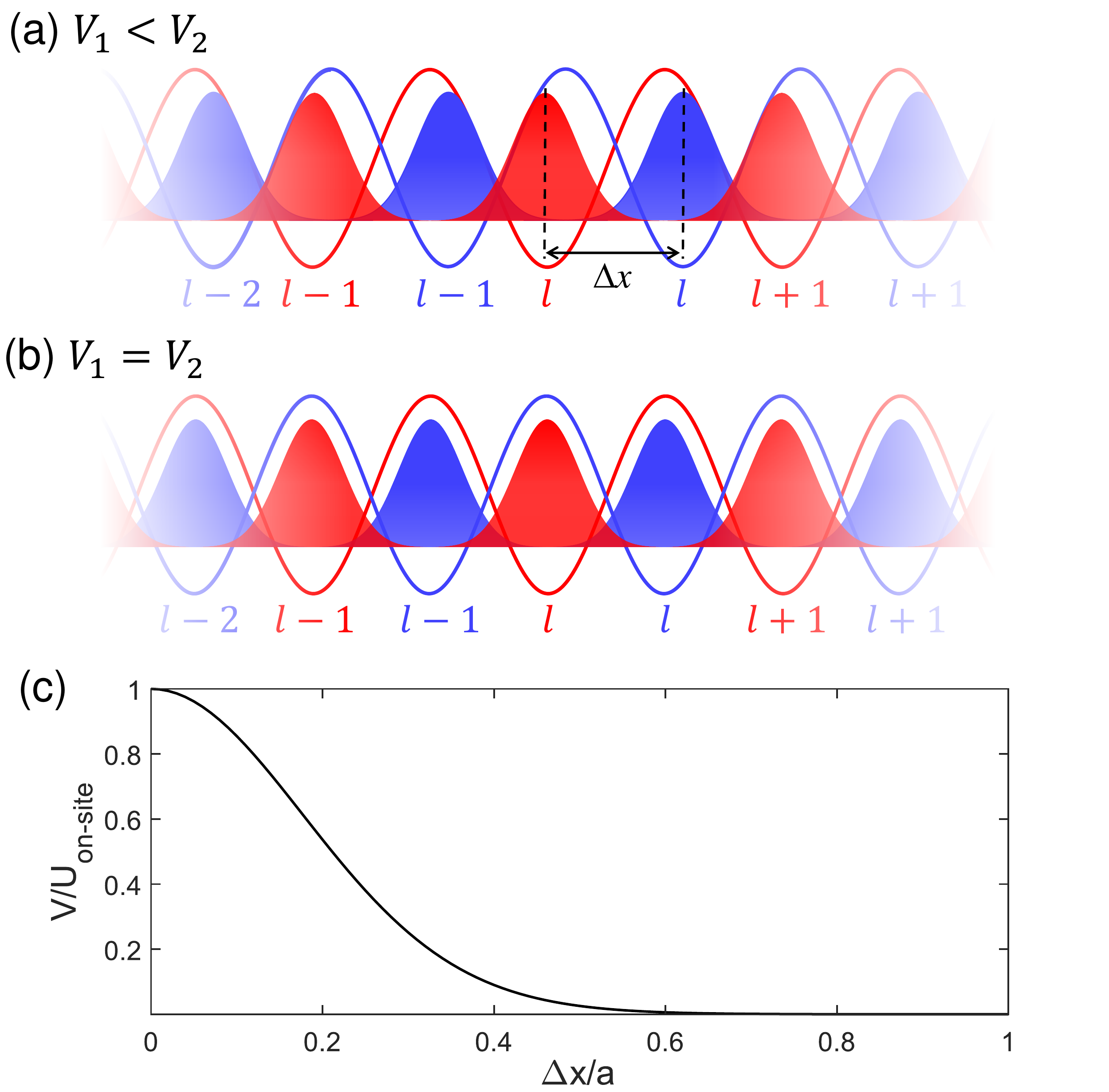}
  \caption{\label{fig:Wannier}
Depict for the intra-species interaction in terms of the highly localized Wannier functions.
Blue and red color respectively indicates the spin-up and spin-down particle and its associated trapping lattice.
The interaction depends on the overlaps of the two Wannier function, and we show two cases as an schematic illustration: (a) $V_1<V_2$, corresponding to $1/2<\Delta x<1$, and (b) $V_1=V_2$, corresponding to $\Delta x =1/2$.
(c) shows the strength of interaction (in the unit of on-site interaction, which is the strongest) as a function of relative distance $\Delta x$ between two Wannier functions.
This result is based on the numerical calculation of the Wannier function of the lowest band with $\mathcal{V}_\sigma ^{\left( 0 \right)} \approx 13 E_R$.
 }
\end{figure}
With tight-binding approximation, one can expand the field operators in the 1D single-particle Wannier basis
\begin{equation}
{{\hat \Psi }_\sigma }\left( x \right) = \sum\limits_l {w_{\sigma}\left( {{x} - {{x}_l}} \right){{\hat c}_{\sigma ,l}}} ,
\end{equation}
where ${w_{\sigma}\left( x-x_l \right)}$ is the Wannier function of spin-${\sigma}$ particle localized around $ x_l $.
Finally, we obtain
\begin{equation}
\hat H =  - \sum\limits_{i,j,\sigma  =  \uparrow , \downarrow } {J_{i,j}^\sigma \hat c_{\sigma ,i}^\dag {{\hat c}_{\sigma ,j}}}  + \sum\limits_{i,j,k,l} {{V_{i,j,k,l}}\hat c_{ \uparrow ,i}^\dag \hat c_{ \downarrow ,j}^\dag {{\hat c}_{ \uparrow ,k}}{{\hat c}_{ \uparrow ,l}}}
\end{equation}
where the tunneling and interaction energy are
\begin{eqnarray}
\label{eqn:appendix_wannier_integral}
J_{i,j}^\sigma  = &&-\int {dx\;{w_\sigma }^*\left( {{x-x_i}} \right)} \nonumber \\
&&\times\left[ { - \frac{{{\hbar ^2}}}{{2m}}{\nabla ^2} + \mathcal{V}_\sigma ^{{\rm{lat}}}\left( x \right)} \right]{w_\sigma }\left( {{x-x_j}} \right) ;\nonumber \\
{V_{i,j,k,l}} = &&\int {dx\;{w_ \uparrow }^*\left( {{x-x_i}} \right){w_ \downarrow }^*\left( {{x-x_j}} \right)}  \nonumber \\
&& \times {w_ \uparrow }\left( {{x-x_k}} \right){w_ \downarrow }\left( {{x-x_l}} \right).
\end{eqnarray}
The strength of inter-species interaction depends on the overlap of Wannier functions, which is affected by their relative distance.
The more closer they are, the stronger they interact, as depicted in Fig.~\ref{fig:Wannier} (a-b).
Thus, we are able to control the inter-species interaction via tunning the relative shift of the two lattices.
The tunneling strength, on the other hand, may be engineered through the spin-dependent ac Stark shift \citep{PhysRevA.70.033603}.
The strengthes of two NN interactions mentioned in main text read as
\begin{eqnarray}
{V_1} &=& \int {dx\;|{w_ \uparrow }\left( {{x - x_l}} \right){|^2}|{w_ \downarrow }\left( {{x - x_l}} \right){|^2}} ; \nonumber \\
{V_2}&=& \int {dx\;|{w_ \uparrow }\left( {{x - x_l}} \right){|^2}|{w_ \downarrow }\left( {{x - x_{l+1}}} \right){|^2}} .
\end{eqnarray}
Then we would like to estimate the strength of the NN interaction $V_{1,2}$.
%
Generally, the NN interaction is omitted for deep optical lattice, as it is very weak compared with the on-site interaction.
However, the NN interaction considered in our model is quite different.
If the relative phase between the two lattices is $\phi=\pi/4$ in Eq.~\eqref{eqn:spin_dependent_optical_potential}, the distance between two nearest Wannier functions of different species is $\Delta x=a/2$.
In this condition, the integral of interaction term in Eq.~\eqref{eqn:appendix_wannier_integral} gives $V_1=V_2$, corresponding to the case in our model.
We numerically \citep{krutitsky2016ultracold}  calculate the integral Eq.~\eqref{eqn:appendix_wannier_integral}for different relative distance $\Delta x$ in Fig.~\ref{fig:Wannier} (c).
We find the NN interaction strength is roughly about $V_1=V_2\approx 0.025U_{\rm{on-site}}$ when $\mathcal{V}_\sigma ^{\left( 0 \right)} \approx 13 E_R$.
Here we use the on-site interaction $U_{\rm{on-site}}$ for comparison.
Usually, the ratio between on-site interaction $U$ and NN tunneling $J$ in optical lattice is in the magnitude of $U_{\rm{on-site}}/J_{\rm{NN}}\approx 36$ when $\mathcal{V}_\sigma ^{\left( 0 \right)} \approx 13 E_R$ for $^{87}\mathrm{Rb}$ \citep{greiner2002quantum}.
Thus, the NN interaction strength is roughly about $V_1=V_2\approx 0.9J$, which is quite considerable.
This estimation is in agreement with the result in Ref.~\citep{PhysRevA.94.063632}, in which similar type of interaction is considered.
With the help of Feshbach resonance, it is possible to increase the s-wave scattering length and produce the desirable strong NN interaction.
As shown in Fig.~\ref{fig:Wannier} (c), the integral of interaction term in Eq.~\eqref{eqn:appendix_wannier_integral} decay fast with the increase of distance.
The interaction strength for $\Delta x=a$ is about $10^{-4}U_{\rm{on-site}} \ll  V_{1,2}$.
Therefore, we may keep only the two NN inter-species interaction terms in our model \eqref{eq:Ham}.

\section{Effect of imbalanced interaction on Zak phase}
\label{appendix:imbalance}
\begin{figure}
  \includegraphics[width = \columnwidth ]{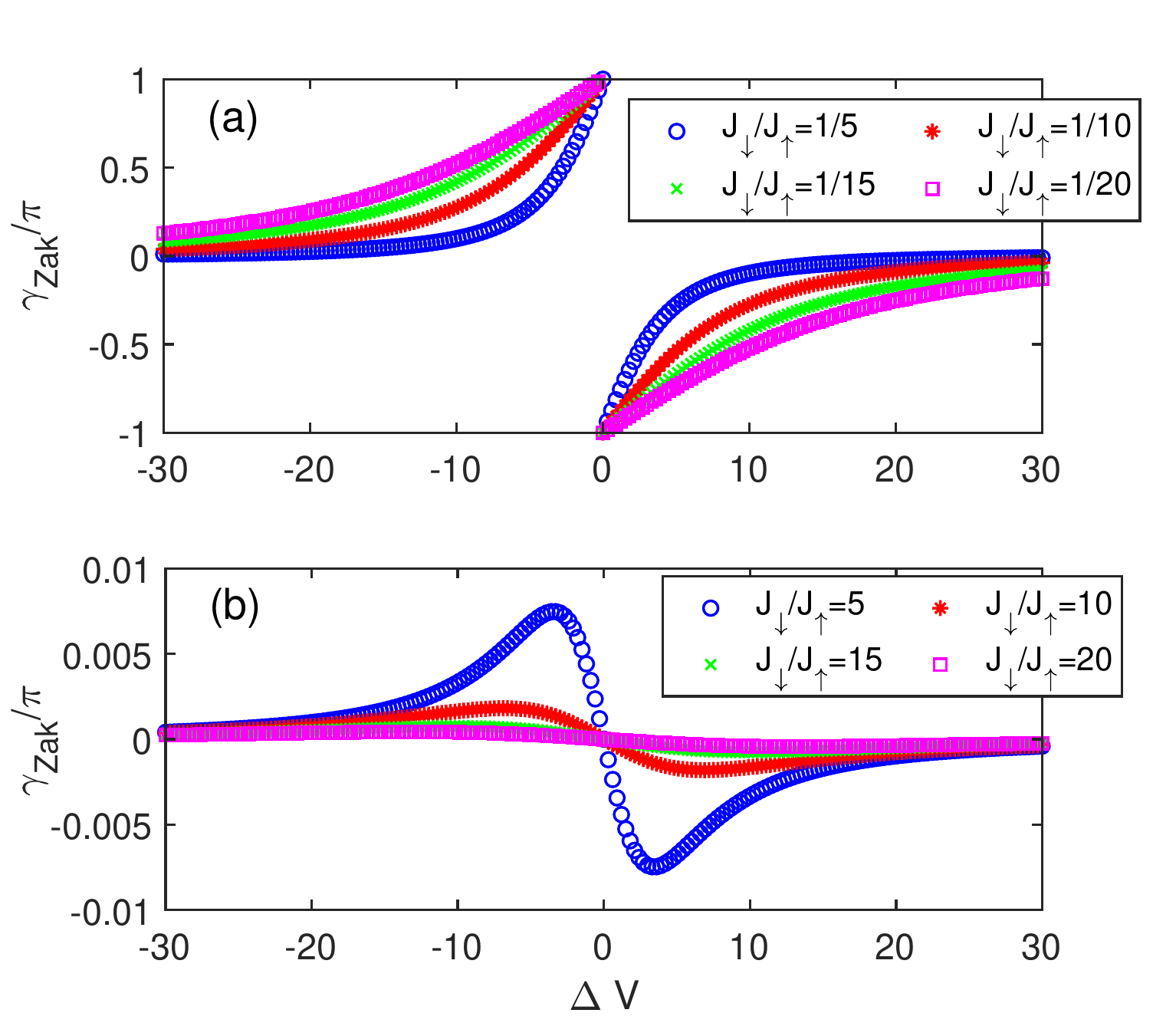}
  \caption{\label{fig:appendix_ZakPhase_imbalance}
 The Zak phase of upper isolated band versus different imbalanced interaction $\Delta V$.
 Here $V_1=V_0+\Delta V$ and $V_2=V_0-\Delta V$ with $V_0=100$;
 (a) and (b) show the results of $J_\downarrow<J_\uparrow$ and $J_\downarrow>J_\uparrow$ respectively.
 We fix $ J_\downarrow=1$ in (a) and $J_\uparrow=1$ in (b).
 For simplicity, we only present the results of upper band.
 }
\end{figure}

We briefly discuss if the two interaction is not equal $V_1 \neq V_2$.
In this case, the c.m. inversion symmetry would be broken, and the relation in Eq.~\eqref{eqn:cm_inversion_sym} does not hold.
Hence, the Zak phase is not quantized to 0 or $\pi$ mode $2\pi$, and will be affected by parameters.
We calculate the Zak phase for different imbalance $\Delta V = (V_1-V_2)/2$ in Fig.~\ref{fig:appendix_ZakPhase_imbalance}.
One can note the non-zero Zak phase will decrease if the imbalance $\Delta V$ increases.
This also explains why the Zak phase changes sharply in Fig.~\ref{fig:Pumping_Scheme} (g), since the imbalance of interaction strength is large for most time.
If the imbalance is small $|\Delta V/J_{1,2}|\ll 1$, the Zak phase would still be close to 0 or $\pi$.
It can be also noted that the effect of imbalance of the interaction in our model is in analogy to imbalance of on-site potential of sublattice in SSH model, in which the chiral symmetry and inversion symmetry would be broken by this imbalance.
To some extents, the considerations on such imbalance in these two models share some similarities.
\begin{figure}
  \includegraphics[width = \columnwidth ]{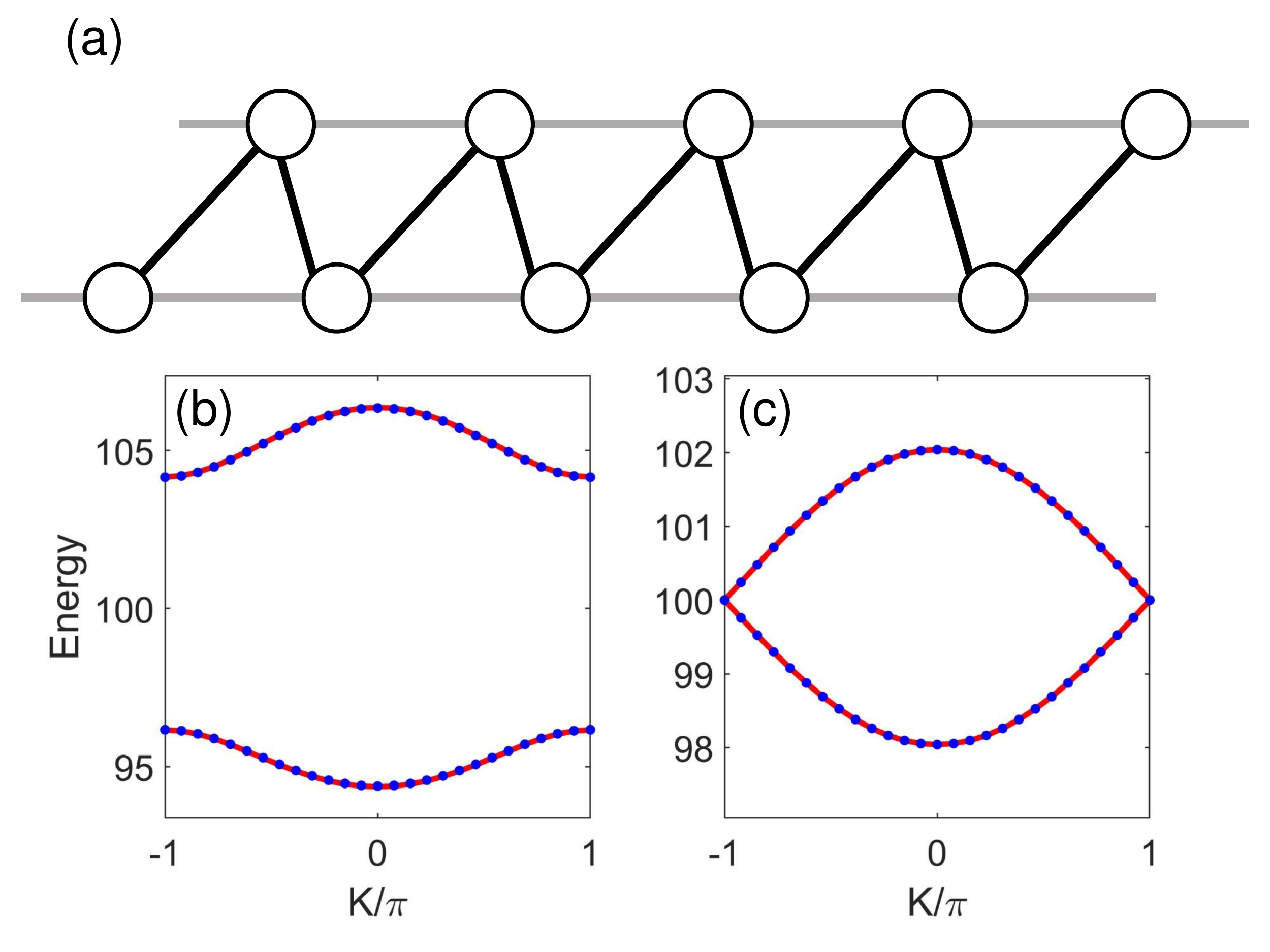}
  \caption{\label{fig:Eff}
 (a) Lattice geometry of the effective model Eq.~\ref{eq:Effective_Model}.
 Black and grey lines represent first- and second-order effective tunneling of bound states.
 (b-c) Comparison between original model and the effective single-particle model.
 Blue dots are calculated numerically from original model, and red lines are from effective model.
 We set $J_\downarrow=1,J_\uparrow=5$ for (b),  $J_\downarrow=J_\uparrow=1$ for (c), and $V_1=V_2=100$ for both.
 }
\end{figure}

\section{Effective model for bound states}\label{appendix:eff_model}
Taking the tunneling as perturbation, one can obtain the effective single-particle model describing the bound states.
For simplicity, we only consider the condition $V_1=V_2=V$.
The unperturbed Hamiltonian reads
\begin{equation}
\label{eq:Ham0}
{{\hat H}_0} = V\sum\limits_l {\left( {{{\hat n}_{\downarrow,l}}{{\hat n}_{\uparrow,l}} +{{\hat n}_{\downarrow,l + 1}}{{\hat n}_{\uparrow,l}}} \right)},
\end{equation}
and the perturbative term is
\begin{equation}
\hat H_J = -\sum\limits_{l,\sigma  =  \uparrow , \downarrow } {\left( {{J_\sigma }\hat c_{\sigma ,l}^\dag {{\hat c}_{\sigma ,l + 1}} + H.c.} \right)}
\end{equation}
Aside from the eigenstates that two particles are separated, one can find two kinds of bound states $|{d_{A,l}}\rangle  = c_{ \downarrow ,l}^\dag c_{ \uparrow ,l}^\dag |0\rangle = |l_\downarrow, l_\uparrow\rangle$ and $|{d_{B,l}}\rangle  = c_{ \downarrow ,l+1}^\dag c_{ \uparrow , l }^\dag |0\rangle = |(l+1)_\downarrow, l_\uparrow\rangle$ with degenerate eigenenergy $E_0=V$.
By applying the degenerate perturbation theory up to second order \cite{JPC1977Takahashi},
\begin{equation}
{\hat H_{{\rm{eff}}}}{\rm{ }} = E_0\hat P + \hat P \hat H_J \hat P +  \hat P \hat H_J \hat S \hat H_J \hat P,
\end{equation}
where $\hat P = \sum_l {(|{d_{A,l}}\rangle \langle {d_{A,l}}| + |{d_{B,l}}\rangle \langle {d_{B,l}}|)}$ is the projector onto the subspace spanned by unperturbed bound states, and $\hat S = (\mathbf{1}-\hat{P}) /V$ is a projector onto the orthogonal component of $\hat{P}$.
Consequently, written in the form of particle operators, one obtains
\begin{eqnarray}
\label{eq:Effective_Model}
{{\hat H}_{{\rm{eff}}}} &= &\left( {V + \frac{{J_\downarrow^2 + J_\uparrow^2}}{V}} \right)\sum\limits_l {\left( {\hat d_{A,l}^\dag {{\hat d}_{A,l}} + \hat d_{B,l}^\dag {{\hat d}_{B,l}}} \right)}   \nonumber \\
&& - \sum\limits_l {\left( {{J_\downarrow}\hat d_{A,l}^\dag {{\hat d}_{B,l}} + {J_\uparrow}\hat d_{B,l}^\dag {{\hat d}_{A,l+ 1}} + H.c.} \right)}   \\
&& + \frac{{{J_\downarrow}{J_\uparrow}}}{V}\sum\limits_i {\left( {\hat d_{A,l}^\dag {{\hat d}_{A,l + 1}} + \hat d_{B,l}^\dag {{\hat d}_{B,l + 1}} + H.c.} \right)} . \nonumber
\end{eqnarray}
The first row in the above effective Hamiltonian is the homogeneous on-site energy.
Second and third rows respectively correspond to dimerized NN tunneling and homogeneous next-nearest-neighbor (NNN) tunneling.
 This effective Hamiltonian indicates the zig-zag geometry, see Fig.~\ref{fig:Eff} (a).
The Fourier transformation yields
\begin{equation}
h\left( k \right) = {d_0}\left( k \right)I + {d_x}\left( k \right){\sigma _x} + {d_y}\left( k \right){\sigma _y},
\end{equation}
in which ${\sigma}_i$ is the Pauli matrices, and ${d_0}\left( k \right) = V + \left( {J_\downarrow^2 + J_\uparrow^2} \right)/V + {J_\downarrow}{J_\uparrow}\cos \left( k \right)/V$, ${d_x}\left( k \right) =   - {J_\uparrow} - {J_\downarrow}\cos \left( k \right)$, ${d_y}\left( k \right) =  - {J_\downarrow}\sin \left( k \right) $.
It can be verified that the eigenenergy of this effective model coincides well with the original model up to the order of $O[(J_\downarrow+J_\uparrow)^3/V^2]$, as compared in Fig.~\ref{fig:Eff} (b-c).
There is the inversion symmetry $\sigma_x h(k) \sigma_x=h(-k)$, and the Zak phase is quantized.
In addition, it can be found that the inversion symmetry will be preserved for arbitrary order of perturbation.
Such kind of property is due to the inversion-symmetric form of interaction, and the tunneling does not break this symmetry.

\section{Asymmetric open boundary condition}
\label{appendix:OBC}
\begin{figure}[htp]
  \includegraphics[width = \columnwidth ]{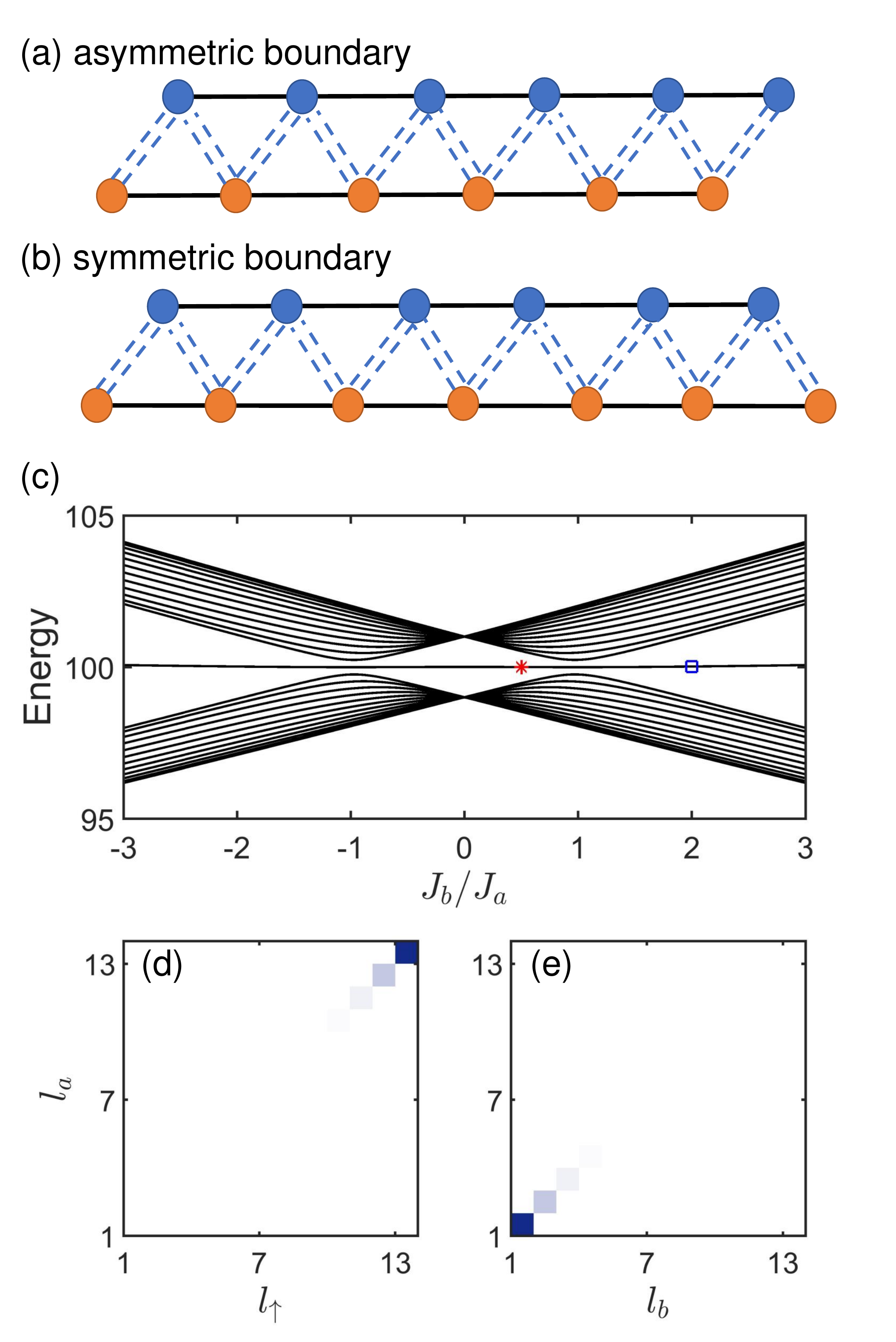}
  \caption{\label{fig:antiOBC}
 (a-b) Lattice geometry of the asymmetric and symmetric boundary.
 Blue-dashed lines indicate the interaction.
 (c) Eigenenergies of bound states versus $J_\uparrow/J_\downarrow$ under the asymmetric-open boundary condition, with same parameters as Fig.~\ref{fig:EdgeBSs} (a).
 (d-e) The correlation matrices $\Gamma_{{l,l'}}  = \langle {n_{a,l}}{n_{{b,l' }}}\rangle$ of the eigenstates corresponds to red star and blue square in (c).
 }
\end{figure}

The schematic diagram of asymmetric and symmetric boundary are shown in Fig.~\ref{fig:antiOBC} (a) and (b).
We calculate the spectrum and in-gap states with asymmetric boundary, see Fig.~\ref{fig:antiOBC} (c-e).
There is always a in-gap state, but the occupations on the edge are different.
This can be understood from the effective model in Appendix.~\ref{appendix:eff_model}.
With asymmetric boundary, the effective lattice misses one site on the edge, and thus the two edges are in different dimerization.
No matter how the ratio of $|J_\uparrow/J_\downarrow|$ changes, there will always a edge bound-state if the lattice is long enough, and the bulk-edge correspondence fails.
The difference of the two kinds of termination as well as the results is because that the bulk-edge correspondence should respect the symmetry that protects the topological properties \citep{PhysRevB.97.115143,PhysRevB.95.035421}.
In our model, the protecting symmetry is the c.m. inversion symmetry, which is broken by the asymmetric termination of edge.
\bibliography{Main_topo_two_body}

\begin{thebibliography}{67}%
\makeatletter
\providecommand \@ifxundefined [1]{%
 \@ifx{#1\undefined}
}%
\providecommand \@ifnum [1]{%
 \ifnum #1\expandafter \@firstoftwo
 \else \expandafter \@secondoftwo
 \fi
}%
\providecommand \@ifx [1]{%
 \ifx #1\expandafter \@firstoftwo
 \else \expandafter \@secondoftwo
 \fi
}%
\providecommand \natexlab [1]{#1}%
\providecommand \enquote  [1]{``#1''}%
\providecommand \bibnamefont  [1]{#1}%
\providecommand \bibfnamefont [1]{#1}%
\providecommand \citenamefont [1]{#1}%
\providecommand \href@noop [0]{\@secondoftwo}%
\providecommand \href [0]{\begingroup \@sanitize@url \@href}%
\providecommand \@href[1]{\@@startlink{#1}\@@href}%
\providecommand \@@href[1]{\endgroup#1\@@endlink}%
\providecommand \@sanitize@url [0]{\catcode `\\12\catcode `\$12\catcode
  `\&12\catcode `\#12\catcode `\^12\catcode `\_12\catcode `\%12\relax}%
\providecommand \@@startlink[1]{}%
\providecommand \@@endlink[0]{}%
\providecommand \url  [0]{\begingroup\@sanitize@url \@url }%
\providecommand \@url [1]{\endgroup\@href {#1}{\urlprefix }}%
\providecommand \urlprefix  [0]{URL }%
\providecommand \Eprint [0]{\href }%
\providecommand \doibase [0]{http://dx.doi.org/}%
\providecommand \selectlanguage [0]{\@gobble}%
\providecommand \bibinfo  [0]{\@secondoftwo}%
\providecommand \bibfield  [0]{\@secondoftwo}%
\providecommand \translation [1]{[#1]}%
\providecommand \BibitemOpen [0]{}%
\providecommand \bibitemStop [0]{}%
\providecommand \bibitemNoStop [0]{.\EOS\space}%
\providecommand \EOS [0]{\spacefactor3000\relax}%
\providecommand \BibitemShut  [1]{\csname bibitem#1\endcsname}%
\let\auto@bib@innerbib\@empty
\bibitem [{\citenamefont {Thouless}\ \emph {et~al.}(1982)\citenamefont
  {Thouless}, \citenamefont {Kohmoto}, \citenamefont {Nightingale},\ and\
  \citenamefont {den Nijs}}]{PhysRevLett.49.405}%
  \BibitemOpen
  \bibfield  {author} {\bibinfo {author} {\bibfnamefont {D.~J.}\ \bibnamefont
  {Thouless}}, \bibinfo {author} {\bibfnamefont {M.}~\bibnamefont {Kohmoto}},
  \bibinfo {author} {\bibfnamefont {M.~P.}\ \bibnamefont {Nightingale}}, \ and\
  \bibinfo {author} {\bibfnamefont {M.}~\bibnamefont {den Nijs}},\ }\bibfield
  {title} {\enquote {\bibinfo {title} {Quantized hall conductance in a
  two-dimensional periodic potential},}\ }\href {\doibase
  10.1103/PhysRevLett.49.405} {\bibfield  {journal} {\bibinfo  {journal} {Phys.
  Rev. Lett.}\ }\textbf {\bibinfo {volume} {49}},\ \bibinfo {pages} {405--408}
  (\bibinfo {year} {1982})}\BibitemShut {NoStop}%
\bibitem [{\citenamefont {Thouless}(1983)}]{PhysRevB.27.6083}%
  \BibitemOpen
  \bibfield  {author} {\bibinfo {author} {\bibfnamefont {D.~J.}\ \bibnamefont
  {Thouless}},\ }\bibfield  {title} {\enquote {\bibinfo {title} {Quantization
  of particle transport},}\ }\href {\doibase 10.1103/PhysRevB.27.6083}
  {\bibfield  {journal} {\bibinfo  {journal} {Phys. Rev. B}\ }\textbf {\bibinfo
  {volume} {27}},\ \bibinfo {pages} {6083--6087} (\bibinfo {year}
  {1983})}\BibitemShut {NoStop}%
\bibitem [{\citenamefont {Hasan}\ and\ \citenamefont
  {Kane}(2010)}]{RevModPhys.82.3045}%
  \BibitemOpen
  \bibfield  {author} {\bibinfo {author} {\bibfnamefont {M.~Z.}\ \bibnamefont
  {Hasan}}\ and\ \bibinfo {author} {\bibfnamefont {C.~L.}\ \bibnamefont
  {Kane}},\ }\bibfield  {title} {\enquote {\bibinfo {title} {Colloquium:
  Topological insulators},}\ }\href {\doibase 10.1103/RevModPhys.82.3045}
  {\bibfield  {journal} {\bibinfo  {journal} {Rev. Mod. Phys.}\ }\textbf
  {\bibinfo {volume} {82}},\ \bibinfo {pages} {3045--3067} (\bibinfo {year}
  {2010})}\BibitemShut {NoStop}%
\bibitem [{\citenamefont {Kane}\ and\ \citenamefont
  {Mele}(2005)}]{PhysRevLett.95.226801}%
  \BibitemOpen
  \bibfield  {author} {\bibinfo {author} {\bibfnamefont {C.~L.}\ \bibnamefont
  {Kane}}\ and\ \bibinfo {author} {\bibfnamefont {E.~J.}\ \bibnamefont
  {Mele}},\ }\bibfield  {title} {\enquote {\bibinfo {title} {Quantum spin hall
  effect in graphene},}\ }\href {\doibase 10.1103/PhysRevLett.95.226801}
  {\bibfield  {journal} {\bibinfo  {journal} {Phys. Rev. Lett.}\ }\textbf
  {\bibinfo {volume} {95}},\ \bibinfo {pages} {226801} (\bibinfo {year}
  {2005})}\BibitemShut {NoStop}%
\bibitem [{\citenamefont {Haldane}(1988)}]{PhysRevLett.61.2015}%
  \BibitemOpen
  \bibfield  {author} {\bibinfo {author} {\bibfnamefont {F.~D.~M.}\
  \bibnamefont {Haldane}},\ }\bibfield  {title} {\enquote {\bibinfo {title}
  {Model for a quantum hall effect without landau levels: Condensed-matter
  realization of the "parity anomaly"},}\ }\href {\doibase
  10.1103/PhysRevLett.61.2015} {\bibfield  {journal} {\bibinfo  {journal}
  {Phys. Rev. Lett.}\ }\textbf {\bibinfo {volume} {61}},\ \bibinfo {pages}
  {2015--2018} (\bibinfo {year} {1988})}\BibitemShut {NoStop}%
\bibitem [{\citenamefont {Bernevig}\ and\ \citenamefont
  {Zhang}(2006)}]{PhysRevLett.96.106802}%
  \BibitemOpen
  \bibfield  {author} {\bibinfo {author} {\bibfnamefont {B.~A.}\ \bibnamefont
  {Bernevig}}\ and\ \bibinfo {author} {\bibfnamefont {S.-C.}\ \bibnamefont
  {Zhang}},\ }\bibfield  {title} {\enquote {\bibinfo {title} {Quantum spin hall
  effect},}\ }\href {\doibase 10.1103/PhysRevLett.96.106802} {\bibfield
  {journal} {\bibinfo  {journal} {Phys. Rev. Lett.}\ }\textbf {\bibinfo
  {volume} {96}},\ \bibinfo {pages} {106802} (\bibinfo {year}
  {2006})}\BibitemShut {NoStop}%
\bibitem [{\citenamefont {Bernevig}\ \emph {et~al.}(2006)\citenamefont
  {Bernevig}, \citenamefont {Hughes},\ and\ \citenamefont
  {Zhang}}]{bernevig2006quantum}%
  \BibitemOpen
  \bibfield  {author} {\bibinfo {author} {\bibfnamefont {B.~A.}\ \bibnamefont
  {Bernevig}}, \bibinfo {author} {\bibfnamefont {T.~L.}\ \bibnamefont
  {Hughes}}, \ and\ \bibinfo {author} {\bibfnamefont {S.-C.}\ \bibnamefont
  {Zhang}},\ }\bibfield  {title} {\enquote {\bibinfo {title} {Quantum spin hall
  effect and topological phase transition in hgte quantum wells},}\ }\href
  {\doibase 10.1126/science.1133734} {\bibfield  {journal} {\bibinfo  {journal}
  {Science}\ }\textbf {\bibinfo {volume} {314}},\ \bibinfo {pages} {1757--1761}
  (\bibinfo {year} {2006})}\BibitemShut {NoStop}%
\bibitem [{\citenamefont {Lohse}\ \emph {et~al.}(2016)\citenamefont {Lohse},
  \citenamefont {Schweizer}, \citenamefont {Zilberberg}, \citenamefont
  {Aidelsburger},\ and\ \citenamefont {Bloch}}]{lohse2016thouless}%
  \BibitemOpen
  \bibfield  {author} {\bibinfo {author} {\bibfnamefont {M.}~\bibnamefont
  {Lohse}}, \bibinfo {author} {\bibfnamefont {C.}~\bibnamefont {Schweizer}},
  \bibinfo {author} {\bibfnamefont {O.}~\bibnamefont {Zilberberg}}, \bibinfo
  {author} {\bibfnamefont {M.}~\bibnamefont {Aidelsburger}}, \ and\ \bibinfo
  {author} {\bibfnamefont {I.}~\bibnamefont {Bloch}},\ }\bibfield  {title}
  {\enquote {\bibinfo {title} {A thouless quantum pump with ultracold bosonic
  atoms in an optical superlattice},}\ }\href {\doibase 10.1038/NPHYS3584}
  {\bibfield  {journal} {\bibinfo  {journal} {Nature Physics}\ }\textbf
  {\bibinfo {volume} {12}},\ \bibinfo {pages} {350} (\bibinfo {year}
  {2016})}\BibitemShut {NoStop}%
\bibitem [{\citenamefont {Nakajima}\ \emph {et~al.}(2016)\citenamefont
  {Nakajima}, \citenamefont {Tomita}, \citenamefont {Taie}, \citenamefont
  {Ichinose}, \citenamefont {Ozawa}, \citenamefont {Wang}, \citenamefont
  {Troyer},\ and\ \citenamefont {Takahashi}}]{nakajima2016topological}%
  \BibitemOpen
  \bibfield  {author} {\bibinfo {author} {\bibfnamefont {S.}~\bibnamefont
  {Nakajima}}, \bibinfo {author} {\bibfnamefont {T.}~\bibnamefont {Tomita}},
  \bibinfo {author} {\bibfnamefont {S.}~\bibnamefont {Taie}}, \bibinfo {author}
  {\bibfnamefont {T.}~\bibnamefont {Ichinose}}, \bibinfo {author}
  {\bibfnamefont {H.}~\bibnamefont {Ozawa}}, \bibinfo {author} {\bibfnamefont
  {L.}~\bibnamefont {Wang}}, \bibinfo {author} {\bibfnamefont {M.}~\bibnamefont
  {Troyer}}, \ and\ \bibinfo {author} {\bibfnamefont {Y.}~\bibnamefont
  {Takahashi}},\ }\bibfield  {title} {\enquote {\bibinfo {title} {Topological
  thouless pumping of ultracold fermions},}\ }\href {\doibase
  10.1038/NPHYS3622} {\bibfield  {journal} {\bibinfo  {journal} {Nature
  Physics}\ }\textbf {\bibinfo {volume} {12}},\ \bibinfo {pages} {296}
  (\bibinfo {year} {2016})}\BibitemShut {NoStop}%
\bibitem [{\citenamefont {Aidelsburger}\ \emph {et~al.}(2013)\citenamefont
  {Aidelsburger}, \citenamefont {Atala}, \citenamefont {Lohse}, \citenamefont
  {Barreiro}, \citenamefont {Paredes},\ and\ \citenamefont
  {Bloch}}]{PhysRevLett.111.185301}%
  \BibitemOpen
  \bibfield  {author} {\bibinfo {author} {\bibfnamefont {M.}~\bibnamefont
  {Aidelsburger}}, \bibinfo {author} {\bibfnamefont {M.}~\bibnamefont {Atala}},
  \bibinfo {author} {\bibfnamefont {M.}~\bibnamefont {Lohse}}, \bibinfo
  {author} {\bibfnamefont {J.~T.}\ \bibnamefont {Barreiro}}, \bibinfo {author}
  {\bibfnamefont {B.}~\bibnamefont {Paredes}}, \ and\ \bibinfo {author}
  {\bibfnamefont {I.}~\bibnamefont {Bloch}},\ }\bibfield  {title} {\enquote
  {\bibinfo {title} {Realization of the hofstadter hamiltonian with ultracold
  atoms in optical lattices},}\ }\href {\doibase
  10.1103/PhysRevLett.111.185301} {\bibfield  {journal} {\bibinfo  {journal}
  {Phys. Rev. Lett.}\ }\textbf {\bibinfo {volume} {111}},\ \bibinfo {pages}
  {185301} (\bibinfo {year} {2013})}\BibitemShut {NoStop}%
\bibitem [{\citenamefont {Jotzu}\ \emph {et~al.}(2014)\citenamefont {Jotzu},
  \citenamefont {Messer}, \citenamefont {Desbuquois}, \citenamefont {Lebrat},
  \citenamefont {Uehlinger}, \citenamefont {Greif},\ and\ \citenamefont
  {Esslinger}}]{jotzu2014experimental}%
  \BibitemOpen
  \bibfield  {author} {\bibinfo {author} {\bibfnamefont {G.}~\bibnamefont
  {Jotzu}}, \bibinfo {author} {\bibfnamefont {M.}~\bibnamefont {Messer}},
  \bibinfo {author} {\bibfnamefont {R.}~\bibnamefont {Desbuquois}}, \bibinfo
  {author} {\bibfnamefont {M.}~\bibnamefont {Lebrat}}, \bibinfo {author}
  {\bibfnamefont {T.}~\bibnamefont {Uehlinger}}, \bibinfo {author}
  {\bibfnamefont {D.}~\bibnamefont {Greif}}, \ and\ \bibinfo {author}
  {\bibfnamefont {T.}~\bibnamefont {Esslinger}},\ }\bibfield  {title} {\enquote
  {\bibinfo {title} {Experimental realization of the topological haldane model
  with ultracold fermions},}\ }\href {\doibase 10.1038/nature13915} {\bibfield
  {journal} {\bibinfo  {journal} {Nature}\ }\textbf {\bibinfo {volume} {515}},\
  \bibinfo {pages} {237} (\bibinfo {year} {2014})}\BibitemShut {NoStop}%
\bibitem [{\citenamefont {Atala}\ \emph {et~al.}(2013)\citenamefont {Atala},
  \citenamefont {Aidelsburger}, \citenamefont {Barreiro}, \citenamefont
  {Abanin}, \citenamefont {Kitagawa}, \citenamefont {Demler},\ and\
  \citenamefont {Bloch}}]{atala2013}%
  \BibitemOpen
  \bibfield  {author} {\bibinfo {author} {\bibfnamefont {M.}~\bibnamefont
  {Atala}}, \bibinfo {author} {\bibfnamefont {M.}~\bibnamefont {Aidelsburger}},
  \bibinfo {author} {\bibfnamefont {J.~T.}\ \bibnamefont {Barreiro}}, \bibinfo
  {author} {\bibfnamefont {D.}~\bibnamefont {Abanin}}, \bibinfo {author}
  {\bibfnamefont {T.}~\bibnamefont {Kitagawa}}, \bibinfo {author}
  {\bibfnamefont {E.}~\bibnamefont {Demler}}, \ and\ \bibinfo {author}
  {\bibfnamefont {I.}~\bibnamefont {Bloch}},\ }\bibfield  {title} {\enquote
  {\bibinfo {title} {Direct measurement of the zak phase in topological bloch
  bands},}\ }\href {\doibase 10.1038/nphys2790} {\bibfield  {journal} {\bibinfo
   {journal} {Nature Physics}\ }\textbf {\bibinfo {volume} {9}},\ \bibinfo
  {pages} {795} (\bibinfo {year} {2013})}\BibitemShut {NoStop}%
\bibitem [{\citenamefont {Meier}\ \emph {et~al.}(2018)\citenamefont {Meier},
  \citenamefont {An}, \citenamefont {Dauphin}, \citenamefont {Maffei},
  \citenamefont {Massignan}, \citenamefont {Hughes},\ and\ \citenamefont
  {Gadway}}]{meier2018observation}%
  \BibitemOpen
  \bibfield  {author} {\bibinfo {author} {\bibfnamefont {E.~J.}\ \bibnamefont
  {Meier}}, \bibinfo {author} {\bibfnamefont {F.~A.}\ \bibnamefont {An}},
  \bibinfo {author} {\bibfnamefont {A.}~\bibnamefont {Dauphin}}, \bibinfo
  {author} {\bibfnamefont {M.}~\bibnamefont {Maffei}}, \bibinfo {author}
  {\bibfnamefont {P.}~\bibnamefont {Massignan}}, \bibinfo {author}
  {\bibfnamefont {T.~L.}\ \bibnamefont {Hughes}}, \ and\ \bibinfo {author}
  {\bibfnamefont {B.}~\bibnamefont {Gadway}},\ }\bibfield  {title} {\enquote
  {\bibinfo {title} {Observation of the topological anderson insulator in
  disordered atomic wires},}\ }\href {\doibase 10.1126/science.aat3406}
  {\bibfield  {journal} {\bibinfo  {journal} {Science}\ }\textbf {\bibinfo
  {volume} {362}},\ \bibinfo {pages} {929--933} (\bibinfo {year}
  {2018})}\BibitemShut {NoStop}%
\bibitem [{\citenamefont {Sun}\ \emph {et~al.}(2018)\citenamefont {Sun},
  \citenamefont {Yi}, \citenamefont {Wang}, \citenamefont {Zhang},
  \citenamefont {Sanders}, \citenamefont {Xu}, \citenamefont {Wang},
  \citenamefont {Schmiedmayer}, \citenamefont {Deng}, \citenamefont {Liu},
  \citenamefont {Chen},\ and\ \citenamefont {Pan}}]{PhysRevLett.121.250403}%
  \BibitemOpen
  \bibfield  {author} {\bibinfo {author} {\bibfnamefont {W.}~\bibnamefont
  {Sun}}, \bibinfo {author} {\bibfnamefont {C.-R.}\ \bibnamefont {Yi}},
  \bibinfo {author} {\bibfnamefont {B.-Z.}\ \bibnamefont {Wang}}, \bibinfo
  {author} {\bibfnamefont {W.-W.}\ \bibnamefont {Zhang}}, \bibinfo {author}
  {\bibfnamefont {B.~C.}\ \bibnamefont {Sanders}}, \bibinfo {author}
  {\bibfnamefont {X.-T.}\ \bibnamefont {Xu}}, \bibinfo {author} {\bibfnamefont
  {Z.-Y.}\ \bibnamefont {Wang}}, \bibinfo {author} {\bibfnamefont
  {J.}~\bibnamefont {Schmiedmayer}}, \bibinfo {author} {\bibfnamefont
  {Y.}~\bibnamefont {Deng}}, \bibinfo {author} {\bibfnamefont {X.-J.}\
  \bibnamefont {Liu}}, \bibinfo {author} {\bibfnamefont {S.}~\bibnamefont
  {Chen}}, \ and\ \bibinfo {author} {\bibfnamefont {J.-W.}\ \bibnamefont
  {Pan}},\ }\bibfield  {title} {\enquote {\bibinfo {title} {Uncover topology by
  quantum quench dynamics},}\ }\href {\doibase 10.1103/PhysRevLett.121.250403}
  {\bibfield  {journal} {\bibinfo  {journal} {Phys. Rev. Lett.}\ }\textbf
  {\bibinfo {volume} {121}},\ \bibinfo {pages} {250403} (\bibinfo {year}
  {2018})}\BibitemShut {NoStop}%
\bibitem [{\citenamefont {Cheuk}\ \emph {et~al.}(2012)\citenamefont {Cheuk},
  \citenamefont {Sommer}, \citenamefont {Hadzibabic}, \citenamefont {Yefsah},
  \citenamefont {Bakr},\ and\ \citenamefont
  {Zwierlein}}]{PhysRevLett.109.095302}%
  \BibitemOpen
  \bibfield  {author} {\bibinfo {author} {\bibfnamefont {Lawrence~W.}\
  \bibnamefont {Cheuk}}, \bibinfo {author} {\bibfnamefont {Ariel~T.}\
  \bibnamefont {Sommer}}, \bibinfo {author} {\bibfnamefont {Zoran}\
  \bibnamefont {Hadzibabic}}, \bibinfo {author} {\bibfnamefont {Tarik}\
  \bibnamefont {Yefsah}}, \bibinfo {author} {\bibfnamefont {Waseem~S.}\
  \bibnamefont {Bakr}}, \ and\ \bibinfo {author} {\bibfnamefont {Martin~W.}\
  \bibnamefont {Zwierlein}},\ }\bibfield  {title} {\enquote {\bibinfo {title}
  {Spin-injection spectroscopy of a spin-orbit coupled fermi gas},}\ }\href
  {\doibase 10.1103/PhysRevLett.109.095302} {\bibfield  {journal} {\bibinfo
  {journal} {Phys. Rev. Lett.}\ }\textbf {\bibinfo {volume} {109}},\ \bibinfo
  {pages} {095302} (\bibinfo {year} {2012})}\BibitemShut {NoStop}%
\bibitem [{\citenamefont {Lin}\ \emph {et~al.}(2011)\citenamefont {Lin},
  \citenamefont {Jim{\'e}nez-Garc{\'\i}a},\ and\ \citenamefont
  {Spielman}}]{lin2011spin}%
  \BibitemOpen
  \bibfield  {author} {\bibinfo {author} {\bibfnamefont {Y.-J.}\ \bibnamefont
  {Lin}}, \bibinfo {author} {\bibfnamefont {K.}~\bibnamefont
  {Jim{\'e}nez-Garc{\'\i}a}}, \ and\ \bibinfo {author} {\bibfnamefont {I.~B.}\
  \bibnamefont {Spielman}},\ }\bibfield  {title} {\enquote {\bibinfo {title}
  {Spin--orbit-coupled bose--einstein condensates},}\ }\href {\doibase
  10.1038/nature09887} {\bibfield  {journal} {\bibinfo  {journal} {Nature}\
  }\textbf {\bibinfo {volume} {471}},\ \bibinfo {pages} {83} (\bibinfo {year}
  {2011})}\BibitemShut {NoStop}%
\bibitem [{\citenamefont {Wang}\ \emph {et~al.}(2012)\citenamefont {Wang},
  \citenamefont {Yu}, \citenamefont {Fu}, \citenamefont {Miao}, \citenamefont
  {Huang}, \citenamefont {Chai}, \citenamefont {Zhai},\ and\ \citenamefont
  {Zhang}}]{PhysRevLett.109.095301}%
  \BibitemOpen
  \bibfield  {author} {\bibinfo {author} {\bibfnamefont {P.}~\bibnamefont
  {Wang}}, \bibinfo {author} {\bibfnamefont {Z.-Q.}\ \bibnamefont {Yu}},
  \bibinfo {author} {\bibfnamefont {Z.}~\bibnamefont {Fu}}, \bibinfo {author}
  {\bibfnamefont {J.}~\bibnamefont {Miao}}, \bibinfo {author} {\bibfnamefont
  {L.}~\bibnamefont {Huang}}, \bibinfo {author} {\bibfnamefont
  {S.}~\bibnamefont {Chai}}, \bibinfo {author} {\bibfnamefont {H.}~\bibnamefont
  {Zhai}}, \ and\ \bibinfo {author} {\bibfnamefont {J.}~\bibnamefont {Zhang}},\
  }\bibfield  {title} {\enquote {\bibinfo {title} {Spin-orbit coupled
  degenerate fermi gases},}\ }\href {\doibase 10.1103/PhysRevLett.109.095301}
  {\bibfield  {journal} {\bibinfo  {journal} {Phys. Rev. Lett.}\ }\textbf
  {\bibinfo {volume} {109}},\ \bibinfo {pages} {095301} (\bibinfo {year}
  {2012})}\BibitemShut {NoStop}%
\bibitem [{\citenamefont {J\"unemann}\ \emph {et~al.}(2017)\citenamefont
  {J\"unemann}, \citenamefont {Piga}, \citenamefont {Ran}, \citenamefont
  {Lewenstein}, \citenamefont {Rizzi},\ and\ \citenamefont
  {Bermudez}}]{PhysRevX.7.031057}%
  \BibitemOpen
  \bibfield  {author} {\bibinfo {author} {\bibfnamefont {J.}~\bibnamefont
  {J\"unemann}}, \bibinfo {author} {\bibfnamefont {A.}~\bibnamefont {Piga}},
  \bibinfo {author} {\bibfnamefont {S.-J.}\ \bibnamefont {Ran}}, \bibinfo
  {author} {\bibfnamefont {M.}~\bibnamefont {Lewenstein}}, \bibinfo {author}
  {\bibfnamefont {M.}~\bibnamefont {Rizzi}}, \ and\ \bibinfo {author}
  {\bibfnamefont {A.}~\bibnamefont {Bermudez}},\ }\bibfield  {title} {\enquote
  {\bibinfo {title} {Exploring interacting topological insulators with
  ultracold atoms: The synthetic creutz-hubbard model},}\ }\href {\doibase
  10.1103/PhysRevX.7.031057} {\bibfield  {journal} {\bibinfo  {journal} {Phys.
  Rev. X}\ }\textbf {\bibinfo {volume} {7}},\ \bibinfo {pages} {031057}
  (\bibinfo {year} {2017})}\BibitemShut {NoStop}%
\bibitem [{\citenamefont {Cooper}\ \emph {et~al.}(2019)\citenamefont {Cooper},
  \citenamefont {Dalibard},\ and\ \citenamefont
  {Spielman}}]{RevModPhys.91.015005}%
  \BibitemOpen
  \bibfield  {author} {\bibinfo {author} {\bibfnamefont {N.~R.}\ \bibnamefont
  {Cooper}}, \bibinfo {author} {\bibfnamefont {J.}~\bibnamefont {Dalibard}}, \
  and\ \bibinfo {author} {\bibfnamefont {I.~B.}\ \bibnamefont {Spielman}},\
  }\bibfield  {title} {\enquote {\bibinfo {title} {Topological bands for
  ultracold atoms},}\ }\href {\doibase 10.1103/RevModPhys.91.015005} {\bibfield
   {journal} {\bibinfo  {journal} {Rev. Mod. Phys.}\ }\textbf {\bibinfo
  {volume} {91}},\ \bibinfo {pages} {015005} (\bibinfo {year}
  {2019})}\BibitemShut {NoStop}%
\bibitem [{\citenamefont {de~L{\'e}s{\'e}leuc}\ \emph
  {et~al.}(2019)\citenamefont {de~L{\'e}s{\'e}leuc}, \citenamefont {Lienhard},
  \citenamefont {Scholl}, \citenamefont {Barredo}, \citenamefont {Weber},
  \citenamefont {Lang}, \citenamefont {B{\"u}chler}, \citenamefont {Lahaye},\
  and\ \citenamefont {Browaeys}}]{de2019observation}%
  \BibitemOpen
  \bibfield  {author} {\bibinfo {author} {\bibfnamefont {S.}~\bibnamefont
  {de~L{\'e}s{\'e}leuc}}, \bibinfo {author} {\bibfnamefont {V.}~\bibnamefont
  {Lienhard}}, \bibinfo {author} {\bibfnamefont {P.}~\bibnamefont {Scholl}},
  \bibinfo {author} {\bibfnamefont {D.}~\bibnamefont {Barredo}}, \bibinfo
  {author} {\bibfnamefont {S.}~\bibnamefont {Weber}}, \bibinfo {author}
  {\bibfnamefont {N.}~\bibnamefont {Lang}}, \bibinfo {author} {\bibfnamefont
  {H.~P.}\ \bibnamefont {B{\"u}chler}}, \bibinfo {author} {\bibfnamefont
  {T.}~\bibnamefont {Lahaye}}, \ and\ \bibinfo {author} {\bibfnamefont
  {A.}~\bibnamefont {Browaeys}},\ }\bibfield  {title} {\enquote {\bibinfo
  {title} {Observation of a symmetry-protected topological phase of interacting
  bosons with rydberg atoms},}\ }\href {\doibase 10.1126/science.aav9105}
  {\bibfield  {journal} {\bibinfo  {journal} {Science}\ }\textbf {\bibinfo
  {volume} {365}},\ \bibinfo {pages} {775--780} (\bibinfo {year}
  {2019})}\BibitemShut {NoStop}%
\bibitem [{\citenamefont {Niu}\ and\ \citenamefont
  {Thouless}(1984)}]{Niu_1984}%
  \BibitemOpen
  \bibfield  {author} {\bibinfo {author} {\bibfnamefont {Q.}~\bibnamefont
  {Niu}}\ and\ \bibinfo {author} {\bibfnamefont {D.~J.}\ \bibnamefont
  {Thouless}},\ }\bibfield  {title} {\enquote {\bibinfo {title} {Quantised
  adiabatic charge transport in the presence of substrate disorder and
  many-body interaction},}\ }\href {\doibase 10.1088/0305-4470/17/12/016}
  {\bibfield  {journal} {\bibinfo  {journal} {J. Phys. A}\ }\textbf {\bibinfo
  {volume} {17}},\ \bibinfo {pages} {2453} (\bibinfo {year}
  {1984})}\BibitemShut {NoStop}%
\bibitem [{\citenamefont {Ke}\ \emph {et~al.}(2017)\citenamefont {Ke},
  \citenamefont {Qin}, \citenamefont {Kivshar},\ and\ \citenamefont
  {Lee}}]{PhysRevA.95.063630}%
  \BibitemOpen
  \bibfield  {author} {\bibinfo {author} {\bibfnamefont {Y.}~\bibnamefont
  {Ke}}, \bibinfo {author} {\bibfnamefont {X.}~\bibnamefont {Qin}}, \bibinfo
  {author} {\bibfnamefont {Y.~S.}\ \bibnamefont {Kivshar}}, \ and\ \bibinfo
  {author} {\bibfnamefont {C.}~\bibnamefont {Lee}},\ }\bibfield  {title}
  {\enquote {\bibinfo {title} {Multiparticle wannier states and thouless
  pumping of interacting bosons},}\ }\href {\doibase
  10.1103/PhysRevA.95.063630} {\bibfield  {journal} {\bibinfo  {journal} {Phys.
  Rev. A}\ }\textbf {\bibinfo {volume} {95}},\ \bibinfo {pages} {063630}
  (\bibinfo {year} {2017})}\BibitemShut {NoStop}%
\bibitem [{\citenamefont {Qin}\ \emph {et~al.}(2018)\citenamefont {Qin},
  \citenamefont {Mei}, \citenamefont {Ke}, \citenamefont {Zhang},\ and\
  \citenamefont {Lee}}]{Qin_2018}%
  \BibitemOpen
  \bibfield  {author} {\bibinfo {author} {\bibfnamefont {X.}~\bibnamefont
  {Qin}}, \bibinfo {author} {\bibfnamefont {F.}~\bibnamefont {Mei}}, \bibinfo
  {author} {\bibfnamefont {Y.}~\bibnamefont {Ke}}, \bibinfo {author}
  {\bibfnamefont {L.}~\bibnamefont {Zhang}}, \ and\ \bibinfo {author}
  {\bibfnamefont {C.}~\bibnamefont {Lee}},\ }\bibfield  {title} {\enquote
  {\bibinfo {title} {Topological invariant and cotranslational symmetry in
  strongly interacting multi-magnon systems},}\ }\href {\doibase
  10.1088/1367-2630/aa9556} {\bibfield  {journal} {\bibinfo  {journal} {New J.
  Phys.}\ }\textbf {\bibinfo {volume} {20}},\ \bibinfo {pages} {013003}
  (\bibinfo {year} {2018})}\BibitemShut {NoStop}%
\bibitem [{\citenamefont {Piil}\ and\ \citenamefont
  {M\o{}lmer}(2007)}]{PhysRevA.76.023607}%
  \BibitemOpen
  \bibfield  {author} {\bibinfo {author} {\bibfnamefont {R.}~\bibnamefont
  {Piil}}\ and\ \bibinfo {author} {\bibfnamefont {K.}~\bibnamefont
  {M\o{}lmer}},\ }\bibfield  {title} {\enquote {\bibinfo {title} {Tunneling
  couplings in discrete lattices, single-particle band structure, and
  eigenstates of interacting atom pairs},}\ }\href {\doibase
  10.1103/PhysRevA.76.023607} {\bibfield  {journal} {\bibinfo  {journal} {Phys.
  Rev. A}\ }\textbf {\bibinfo {volume} {76}},\ \bibinfo {pages} {023607}
  (\bibinfo {year} {2007})}\BibitemShut {NoStop}%
\bibitem [{\citenamefont {Valiente}\ and\ \citenamefont
  {Petrosyan}(2008)}]{jopB.41.16.161002}%
  \BibitemOpen
  \bibfield  {author} {\bibinfo {author} {\bibfnamefont {M.}~\bibnamefont
  {Valiente}}\ and\ \bibinfo {author} {\bibfnamefont {D.}~\bibnamefont
  {Petrosyan}},\ }\bibfield  {title} {\enquote {\bibinfo {title} {Two-particle
  states in the hubbard model},}\ }\href
  {http://stacks.iop.org/0953-4075/41/i=16/a=161002} {\bibfield  {journal}
  {\bibinfo  {journal} {J. Phys. B: At., Mol. Opt. Phys.}\ }\textbf {\bibinfo
  {volume} {41}},\ \bibinfo {pages} {161002} (\bibinfo {year}
  {2008})}\BibitemShut {NoStop}%
\bibitem [{\citenamefont {Nguenang}\ and\ \citenamefont
  {Flach}(2009)}]{PhysRevA.80.015601}%
  \BibitemOpen
  \bibfield  {author} {\bibinfo {author} {\bibfnamefont {Jean-Pierre}\
  \bibnamefont {Nguenang}}\ and\ \bibinfo {author} {\bibfnamefont
  {S.}~\bibnamefont {Flach}},\ }\bibfield  {title} {\enquote {\bibinfo {title}
  {Fermionic bound states on a one-dimensional lattice},}\ }\href {\doibase
  10.1103/PhysRevA.80.015601} {\bibfield  {journal} {\bibinfo  {journal} {Phys.
  Rev. A}\ }\textbf {\bibinfo {volume} {80}},\ \bibinfo {pages} {015601}
  (\bibinfo {year} {2009})}\BibitemShut {NoStop}%
\bibitem [{\citenamefont {Javanainen}\ \emph {et~al.}(2010)\citenamefont
  {Javanainen}, \citenamefont {Odong},\ and\ \citenamefont
  {Sanders}}]{PhysRevA.81.043609}%
  \BibitemOpen
  \bibfield  {author} {\bibinfo {author} {\bibfnamefont {Juha}\ \bibnamefont
  {Javanainen}}, \bibinfo {author} {\bibfnamefont {Otim}\ \bibnamefont
  {Odong}}, \ and\ \bibinfo {author} {\bibfnamefont {Jerome~C.}\ \bibnamefont
  {Sanders}},\ }\bibfield  {title} {\enquote {\bibinfo {title} {Dimer of two
  bosons in a one-dimensional optical lattice},}\ }\href {\doibase
  10.1103/PhysRevA.81.043609} {\bibfield  {journal} {\bibinfo  {journal} {Phys.
  Rev. A}\ }\textbf {\bibinfo {volume} {81}},\ \bibinfo {pages} {043609}
  (\bibinfo {year} {2010})}\BibitemShut {NoStop}%
\bibitem [{\citenamefont {Winkler}\ \emph {et~al.}(2006)\citenamefont
  {Winkler}, \citenamefont {Thalhammer}, \citenamefont {Lang}, \citenamefont
  {Grimm}, \citenamefont {Denschlag}, \citenamefont {Daley}, \citenamefont
  {Kantian}, \citenamefont {B{\"u}chler},\ and\ \citenamefont
  {Zoller}}]{winkler2006repulsively}%
  \BibitemOpen
  \bibfield  {author} {\bibinfo {author} {\bibfnamefont {K.}~\bibnamefont
  {Winkler}}, \bibinfo {author} {\bibfnamefont {G.}~\bibnamefont {Thalhammer}},
  \bibinfo {author} {\bibfnamefont {F.}~\bibnamefont {Lang}}, \bibinfo {author}
  {\bibfnamefont {R.}~\bibnamefont {Grimm}}, \bibinfo {author} {\bibfnamefont
  {J.~H.}\ \bibnamefont {Denschlag}}, \bibinfo {author} {\bibfnamefont
  {A.}~\bibnamefont {Daley}}, \bibinfo {author} {\bibfnamefont
  {A.}~\bibnamefont {Kantian}}, \bibinfo {author} {\bibfnamefont
  {H.}~\bibnamefont {B{\"u}chler}}, \ and\ \bibinfo {author} {\bibfnamefont
  {P.}~\bibnamefont {Zoller}},\ }\bibfield  {title} {\enquote {\bibinfo {title}
  {Repulsively bound atom pairs in an optical lattice},}\ }\href {\doibase
  10.1038/nature04918} {\bibfield  {journal} {\bibinfo  {journal} {Nature}\
  }\textbf {\bibinfo {volume} {441}},\ \bibinfo {pages} {853} (\bibinfo {year}
  {2006})}\BibitemShut {NoStop}%
\bibitem [{\citenamefont {Di~Liberto}\ \emph {et~al.}(2016)\citenamefont
  {Di~Liberto}, \citenamefont {Recati}, \citenamefont {Carusotto},\ and\
  \citenamefont {Menotti}}]{PhysRevA.94.062704}%
  \BibitemOpen
  \bibfield  {author} {\bibinfo {author} {\bibfnamefont {M.}~\bibnamefont
  {Di~Liberto}}, \bibinfo {author} {\bibfnamefont {A.}~\bibnamefont {Recati}},
  \bibinfo {author} {\bibfnamefont {I.}~\bibnamefont {Carusotto}}, \ and\
  \bibinfo {author} {\bibfnamefont {C.}~\bibnamefont {Menotti}},\ }\bibfield
  {title} {\enquote {\bibinfo {title} {Two-body physics in the
  su-schrieffer-heeger model},}\ }\href {\doibase 10.1103/PhysRevA.94.062704}
  {\bibfield  {journal} {\bibinfo  {journal} {Phys. Rev. A}\ }\textbf {\bibinfo
  {volume} {94}},\ \bibinfo {pages} {062704} (\bibinfo {year}
  {2016})}\BibitemShut {NoStop}%
\bibitem [{\citenamefont {Gorlach}\ and\ \citenamefont
  {Poddubny}(2017)}]{PhysRevA.95.053866}%
  \BibitemOpen
  \bibfield  {author} {\bibinfo {author} {\bibfnamefont {Maxim~A.}\
  \bibnamefont {Gorlach}}\ and\ \bibinfo {author} {\bibfnamefont
  {Alexander~N.}\ \bibnamefont {Poddubny}},\ }\bibfield  {title} {\enquote
  {\bibinfo {title} {Topological edge states of bound photon pairs},}\ }\href
  {\doibase 10.1103/PhysRevA.95.053866} {\bibfield  {journal} {\bibinfo
  {journal} {Phys. Rev. A}\ }\textbf {\bibinfo {volume} {95}},\ \bibinfo
  {pages} {053866} (\bibinfo {year} {2017})}\BibitemShut {NoStop}%
\bibitem [{\citenamefont {Marques}\ and\ \citenamefont
  {Dias}(2017)}]{PhysRevB.95.115443}%
  \BibitemOpen
  \bibfield  {author} {\bibinfo {author} {\bibfnamefont {A.~M.}\ \bibnamefont
  {Marques}}\ and\ \bibinfo {author} {\bibfnamefont {R.~G.}\ \bibnamefont
  {Dias}},\ }\bibfield  {title} {\enquote {\bibinfo {title} {Multihole edge
  states in su-schrieffer-heeger chains with interactions},}\ }\href {\doibase
  10.1103/PhysRevB.95.115443} {\bibfield  {journal} {\bibinfo  {journal} {Phys.
  Rev. B}\ }\textbf {\bibinfo {volume} {95}},\ \bibinfo {pages} {115443}
  (\bibinfo {year} {2017})}\BibitemShut {NoStop}%
\bibitem [{\citenamefont {Marques}\ and\ \citenamefont
  {Dias}(2018)}]{Marques_2018}%
  \BibitemOpen
  \bibfield  {author} {\bibinfo {author} {\bibfnamefont {A.~M.}\ \bibnamefont
  {Marques}}\ and\ \bibinfo {author} {\bibfnamefont {R.~G.}\ \bibnamefont
  {Dias}},\ }\bibfield  {title} {\enquote {\bibinfo {title} {Topological bound
  states in interacting su{\textendash}schrieffer{\textendash}heeger rings},}\
  }\href {\doibase 10.1088/1361-648x/aacd7c} {\bibfield  {journal} {\bibinfo
  {journal} {J. Phys.: Condens. Matter}\ }\textbf {\bibinfo {volume} {30}},\
  \bibinfo {pages} {305601} (\bibinfo {year} {2018})}\BibitemShut {NoStop}%
\bibitem [{\citenamefont {Qin}\ \emph {et~al.}(2017)\citenamefont {Qin},
  \citenamefont {Mei}, \citenamefont {Ke}, \citenamefont {Zhang},\ and\
  \citenamefont {Lee}}]{PhysRevB.96.195134}%
  \BibitemOpen
  \bibfield  {author} {\bibinfo {author} {\bibfnamefont {X.}~\bibnamefont
  {Qin}}, \bibinfo {author} {\bibfnamefont {F.}~\bibnamefont {Mei}}, \bibinfo
  {author} {\bibfnamefont {Y.}~\bibnamefont {Ke}}, \bibinfo {author}
  {\bibfnamefont {L.}~\bibnamefont {Zhang}}, \ and\ \bibinfo {author}
  {\bibfnamefont {C.}~\bibnamefont {Lee}},\ }\bibfield  {title} {\enquote
  {\bibinfo {title} {Topological magnon bound states in periodically modulated
  heisenberg xxz chains},}\ }\href {\doibase 10.1103/PhysRevB.96.195134}
  {\bibfield  {journal} {\bibinfo  {journal} {Phys. Rev. B}\ }\textbf {\bibinfo
  {volume} {96}},\ \bibinfo {pages} {195134} (\bibinfo {year}
  {2017})}\BibitemShut {NoStop}%
\bibitem [{\citenamefont {Salerno}\ \emph {et~al.}(2018)\citenamefont
  {Salerno}, \citenamefont {Di~Liberto}, \citenamefont {Menotti},\ and\
  \citenamefont {Carusotto}}]{PhysRevA.97.013637}%
  \BibitemOpen
  \bibfield  {author} {\bibinfo {author} {\bibfnamefont {G.}~\bibnamefont
  {Salerno}}, \bibinfo {author} {\bibfnamefont {M.}~\bibnamefont {Di~Liberto}},
  \bibinfo {author} {\bibfnamefont {C.}~\bibnamefont {Menotti}}, \ and\
  \bibinfo {author} {\bibfnamefont {I.}~\bibnamefont {Carusotto}},\ }\bibfield
  {title} {\enquote {\bibinfo {title} {Topological two-body bound states in the
  interacting haldane model},}\ }\href {\doibase 10.1103/PhysRevA.97.013637}
  {\bibfield  {journal} {\bibinfo  {journal} {Phys. Rev. A}\ }\textbf {\bibinfo
  {volume} {97}},\ \bibinfo {pages} {013637} (\bibinfo {year}
  {2018})}\BibitemShut {NoStop}%
\bibitem [{\citenamefont {Zhong}\ \emph {et~al.}(2017)\citenamefont {Zhong},
  \citenamefont {Zhou}, \citenamefont {Zhu}, \citenamefont {Ke},\ and\
  \citenamefont {Lee}}]{zhong2017floquet}%
  \BibitemOpen
  \bibfield  {author} {\bibinfo {author} {\bibfnamefont {H.}~\bibnamefont
  {Zhong}}, \bibinfo {author} {\bibfnamefont {Z.}~\bibnamefont {Zhou}},
  \bibinfo {author} {\bibfnamefont {B.}~\bibnamefont {Zhu}}, \bibinfo {author}
  {\bibfnamefont {Y.}~\bibnamefont {Ke}}, \ and\ \bibinfo {author}
  {\bibfnamefont {C.}~\bibnamefont {Lee}},\ }\bibfield  {title} {\enquote
  {\bibinfo {title} {Floquet bound states in a driven two-particle
  bose--hubbard model with an impurity},}\ }\href {\doibase
  10.1088/0256-307X/34/7/070304} {\bibfield  {journal} {\bibinfo  {journal}
  {Chinese Physics Letters}\ }\textbf {\bibinfo {volume} {34}},\ \bibinfo
  {pages} {070304} (\bibinfo {year} {2017})}\BibitemShut {NoStop}%
\bibitem [{\citenamefont {Dobard\ifmmode \check{z}\else
  \v{z}\fi{}i\ifmmode~\acute{c}\else \'{c}\fi{}}\ \emph
  {et~al.}(2015)\citenamefont {Dobard\ifmmode \check{z}\else
  \v{z}\fi{}i\ifmmode~\acute{c}\else \'{c}\fi{}}, \citenamefont
  {Dimitrijevi\ifmmode~\acute{c}\else \'{c}\fi{}},\ and\ \citenamefont
  {Milovanovi\ifmmode~\acute{c}\else \'{c}\fi{}}}]{PhysRevB.91.125424}%
  \BibitemOpen
  \bibfield  {author} {\bibinfo {author} {\bibfnamefont {E.}~\bibnamefont
  {Dobard\ifmmode \check{z}\else \v{z}\fi{}i\ifmmode~\acute{c}\else
  \'{c}\fi{}}}, \bibinfo {author} {\bibfnamefont {M.}~\bibnamefont
  {Dimitrijevi\ifmmode~\acute{c}\else \'{c}\fi{}}}, \ and\ \bibinfo {author}
  {\bibfnamefont {M.~V.}\ \bibnamefont {Milovanovi\ifmmode~\acute{c}\else
  \'{c}\fi{}}},\ }\bibfield  {title} {\enquote {\bibinfo {title} {Generalized
  bloch theorem and topological characterization},}\ }\href {\doibase
  10.1103/PhysRevB.91.125424} {\bibfield  {journal} {\bibinfo  {journal} {Phys.
  Rev. B}\ }\textbf {\bibinfo {volume} {91}},\ \bibinfo {pages} {125424}
  (\bibinfo {year} {2015})}\BibitemShut {NoStop}%
\bibitem [{\citenamefont {Zak}(1989)}]{PhysRevLett.62.2747}%
  \BibitemOpen
  \bibfield  {author} {\bibinfo {author} {\bibfnamefont {J.}~\bibnamefont
  {Zak}},\ }\bibfield  {title} {\enquote {\bibinfo {title} {Berry's phase for
  energy bands in solids},}\ }\href {\doibase 10.1103/PhysRevLett.62.2747}
  {\bibfield  {journal} {\bibinfo  {journal} {Phys. Rev. Lett.}\ }\textbf
  {\bibinfo {volume} {62}},\ \bibinfo {pages} {2747--2750} (\bibinfo {year}
  {1989})}\BibitemShut {NoStop}%
\bibitem [{\citenamefont {Zak}(1982)}]{PhysRevLett.48.359}%
  \BibitemOpen
  \bibfield  {author} {\bibinfo {author} {\bibfnamefont {J.}~\bibnamefont
  {Zak}},\ }\bibfield  {title} {\enquote {\bibinfo {title} {Band center---a
  conserved quantity in solids},}\ }\href {\doibase 10.1103/PhysRevLett.48.359}
  {\bibfield  {journal} {\bibinfo  {journal} {Phys. Rev. Lett.}\ }\textbf
  {\bibinfo {volume} {48}},\ \bibinfo {pages} {359--362} (\bibinfo {year}
  {1982})}\BibitemShut {NoStop}%
\bibitem [{\citenamefont {Resta}(1994)}]{RevModPhys.66.899}%
  \BibitemOpen
  \bibfield  {author} {\bibinfo {author} {\bibfnamefont {Raffaele}\
  \bibnamefont {Resta}},\ }\bibfield  {title} {\enquote {\bibinfo {title}
  {Macroscopic polarization in crystalline dielectrics: the geometric phase
  approach},}\ }\href {\doibase 10.1103/RevModPhys.66.899} {\bibfield
  {journal} {\bibinfo  {journal} {Rev. Mod. Phys.}\ }\textbf {\bibinfo {volume}
  {66}},\ \bibinfo {pages} {899--915} (\bibinfo {year} {1994})}\BibitemShut
  {NoStop}%
\bibitem [{\citenamefont {Benalcazar}\ \emph {et~al.}(2017)\citenamefont
  {Benalcazar}, \citenamefont {Bernevig},\ and\ \citenamefont
  {Hughes}}]{PhysRevB.96.245115}%
  \BibitemOpen
  \bibfield  {author} {\bibinfo {author} {\bibfnamefont {W.~A.}\ \bibnamefont
  {Benalcazar}}, \bibinfo {author} {\bibfnamefont {B.~A.}\ \bibnamefont
  {Bernevig}}, \ and\ \bibinfo {author} {\bibfnamefont {T.~L.}\ \bibnamefont
  {Hughes}},\ }\bibfield  {title} {\enquote {\bibinfo {title} {Electric
  multipole moments, topological multipole moment pumping, and chiral hinge
  states in crystalline insulators},}\ }\href {\doibase
  10.1103/PhysRevB.96.245115} {\bibfield  {journal} {\bibinfo  {journal} {Phys.
  Rev. B}\ }\textbf {\bibinfo {volume} {96}},\ \bibinfo {pages} {245115}
  (\bibinfo {year} {2017})}\BibitemShut {NoStop}%
\bibitem [{\citenamefont {Resta}(1998)}]{PhysRevLett.80.1800}%
  \BibitemOpen
  \bibfield  {author} {\bibinfo {author} {\bibfnamefont {R.}~\bibnamefont
  {Resta}},\ }\bibfield  {title} {\enquote {\bibinfo {title}
  {Quantum-mechanical position operator in extended systems},}\ }\href
  {\doibase 10.1103/PhysRevLett.80.1800} {\bibfield  {journal} {\bibinfo
  {journal} {Phys. Rev. Lett.}\ }\textbf {\bibinfo {volume} {80}},\ \bibinfo
  {pages} {1800--1803} (\bibinfo {year} {1998})}\BibitemShut {NoStop}%
\bibitem [{\citenamefont {Jaksch}\ \emph {et~al.}(1999)\citenamefont {Jaksch},
  \citenamefont {Briegel}, \citenamefont {Cirac}, \citenamefont {Gardiner},\
  and\ \citenamefont {Zoller}}]{PhysRevLett.82.1975}%
  \BibitemOpen
  \bibfield  {author} {\bibinfo {author} {\bibfnamefont {D.}~\bibnamefont
  {Jaksch}}, \bibinfo {author} {\bibfnamefont {H.-J.}\ \bibnamefont {Briegel}},
  \bibinfo {author} {\bibfnamefont {J.~I.}\ \bibnamefont {Cirac}}, \bibinfo
  {author} {\bibfnamefont {C.~W.}\ \bibnamefont {Gardiner}}, \ and\ \bibinfo
  {author} {\bibfnamefont {P.}~\bibnamefont {Zoller}},\ }\bibfield  {title}
  {\enquote {\bibinfo {title} {Entanglement of atoms via cold controlled
  collisions},}\ }\href {\doibase 10.1103/PhysRevLett.82.1975} {\bibfield
  {journal} {\bibinfo  {journal} {Phys. Rev. Lett.}\ }\textbf {\bibinfo
  {volume} {82}},\ \bibinfo {pages} {1975--1978} (\bibinfo {year}
  {1999})}\BibitemShut {NoStop}%
\bibitem [{\citenamefont {Olaf}\ \emph {et~al.}(2003)\citenamefont {Olaf},
  \citenamefont {Markus}, \citenamefont {Artur}, \citenamefont {Tim},
  \citenamefont {H?Nsch},\ and\ \citenamefont {Immanuel}}]{Olaf2003Controlled}%
  \BibitemOpen
  \bibfield  {author} {\bibinfo {author} {\bibfnamefont {M.}~\bibnamefont
  {Olaf}}, \bibinfo {author} {\bibfnamefont {G.}~\bibnamefont {Markus}},
  \bibinfo {author} {\bibfnamefont {W.}~\bibnamefont {Artur}}, \bibinfo
  {author} {\bibfnamefont {R.}~\bibnamefont {Tim}}, \bibinfo {author}
  {\bibfnamefont {T.~W.}\ \bibnamefont {H?Nsch}}, \ and\ \bibinfo {author}
  {\bibfnamefont {B.}~\bibnamefont {Immanuel}},\ }\bibfield  {title} {\enquote
  {\bibinfo {title} {Controlled collisions for multi-particle entanglement of
  optically trapped atoms},}\ }\href {\doibase 10.1038/nature02008} {\bibfield
  {journal} {\bibinfo  {journal} {Nature}\ }\textbf {\bibinfo {volume} {425}},\
  \bibinfo {pages} {937--40} (\bibinfo {year} {2003})}\BibitemShut {NoStop}%
\bibitem [{\citenamefont {Yang}\ \emph {et~al.}(2017)\citenamefont {Yang},
  \citenamefont {Dai}, \citenamefont {Sun}, \citenamefont {Reingruber},
  \citenamefont {Yuan},\ and\ \citenamefont {Pan}}]{PhysRevA.96.011602}%
  \BibitemOpen
  \bibfield  {author} {\bibinfo {author} {\bibfnamefont {B.}~\bibnamefont
  {Yang}}, \bibinfo {author} {\bibfnamefont {H.-N.}\ \bibnamefont {Dai}},
  \bibinfo {author} {\bibfnamefont {H.}~\bibnamefont {Sun}}, \bibinfo {author}
  {\bibfnamefont {A.}~\bibnamefont {Reingruber}}, \bibinfo {author}
  {\bibfnamefont {Z.-S.}\ \bibnamefont {Yuan}}, \ and\ \bibinfo {author}
  {\bibfnamefont {J.-W.}\ \bibnamefont {Pan}},\ }\bibfield  {title} {\enquote
  {\bibinfo {title} {Spin-dependent optical superlattice},}\ }\href {\doibase
  10.1103/PhysRevA.96.011602} {\bibfield  {journal} {\bibinfo  {journal} {Phys.
  Rev. A}\ }\textbf {\bibinfo {volume} {96}},\ \bibinfo {pages} {011602(R)}
  (\bibinfo {year} {2017})}\BibitemShut {NoStop}%
\bibitem [{\citenamefont {Mandel}\ \emph
  {et~al.}(2003{\natexlab{a}})\citenamefont {Mandel}, \citenamefont {Greiner},
  \citenamefont {Widera}, \citenamefont {Rom}, \citenamefont {H\"ansch},\ and\
  \citenamefont {Bloch}}]{PhysRevLett.91.010407}%
  \BibitemOpen
  \bibfield  {author} {\bibinfo {author} {\bibfnamefont {O.}~\bibnamefont
  {Mandel}}, \bibinfo {author} {\bibfnamefont {M.}~\bibnamefont {Greiner}},
  \bibinfo {author} {\bibfnamefont {A.}~\bibnamefont {Widera}}, \bibinfo
  {author} {\bibfnamefont {T.}~\bibnamefont {Rom}}, \bibinfo {author}
  {\bibfnamefont {T.~W.}\ \bibnamefont {H\"ansch}}, \ and\ \bibinfo {author}
  {\bibfnamefont {I.}~\bibnamefont {Bloch}},\ }\bibfield  {title} {\enquote
  {\bibinfo {title} {Coherent transport of neutral atoms in spin-dependent
  optical lattice potentials},}\ }\href {\doibase
  10.1103/PhysRevLett.91.010407} {\bibfield  {journal} {\bibinfo  {journal}
  {Phys. Rev. Lett.}\ }\textbf {\bibinfo {volume} {91}},\ \bibinfo {pages}
  {010407} (\bibinfo {year} {2003}{\natexlab{a}})}\BibitemShut {NoStop}%
\bibitem [{\citenamefont {Holland}\ \emph {et~al.}(2001)\citenamefont
  {Holland}, \citenamefont {Kokkelmans}, \citenamefont {Chiofalo},\ and\
  \citenamefont {Walser}}]{PhysRevLett.87.120406}%
  \BibitemOpen
  \bibfield  {author} {\bibinfo {author} {\bibfnamefont {M.}~\bibnamefont
  {Holland}}, \bibinfo {author} {\bibfnamefont {S.~J. J. M.~F.}\ \bibnamefont
  {Kokkelmans}}, \bibinfo {author} {\bibfnamefont {M.~L.}\ \bibnamefont
  {Chiofalo}}, \ and\ \bibinfo {author} {\bibfnamefont {R.}~\bibnamefont
  {Walser}},\ }\bibfield  {title} {\enquote {\bibinfo {title} {Resonance
  superfluidity in a quantum degenerate fermi gas},}\ }\href {\doibase
  10.1103/PhysRevLett.87.120406} {\bibfield  {journal} {\bibinfo  {journal}
  {Phys. Rev. Lett.}\ }\textbf {\bibinfo {volume} {87}},\ \bibinfo {pages}
  {120406} (\bibinfo {year} {2001})}\BibitemShut {NoStop}%
\bibitem [{\citenamefont {Kariyado}\ and\ \citenamefont
  {Hatsugai}(2013)}]{PhysRevB.88.245126}%
  \BibitemOpen
  \bibfield  {author} {\bibinfo {author} {\bibfnamefont {T.}~\bibnamefont
  {Kariyado}}\ and\ \bibinfo {author} {\bibfnamefont {Y.}~\bibnamefont
  {Hatsugai}},\ }\bibfield  {title} {\enquote {\bibinfo {title}
  {Symmetry-protected quantization and bulk-edge correspondence of massless
  dirac fermions: Application to the fermionic shastry-sutherland model},}\
  }\href {\doibase 10.1103/PhysRevB.88.245126} {\bibfield  {journal} {\bibinfo
  {journal} {Phys. Rev. B}\ }\textbf {\bibinfo {volume} {88}},\ \bibinfo
  {pages} {245126} (\bibinfo {year} {2013})}\BibitemShut {NoStop}%
\bibitem [{\citenamefont {Rhim}\ \emph {et~al.}(2017)\citenamefont {Rhim},
  \citenamefont {Behrends},\ and\ \citenamefont
  {Bardarson}}]{PhysRevB.95.035421}%
  \BibitemOpen
  \bibfield  {author} {\bibinfo {author} {\bibfnamefont {J.-W.}\ \bibnamefont
  {Rhim}}, \bibinfo {author} {\bibfnamefont {J.}~\bibnamefont {Behrends}}, \
  and\ \bibinfo {author} {\bibfnamefont {J.~H.}\ \bibnamefont {Bardarson}},\
  }\bibfield  {title} {\enquote {\bibinfo {title} {Bulk-boundary correspondence
  from the intercellular zak phase},}\ }\href {\doibase
  10.1103/PhysRevB.95.035421} {\bibfield  {journal} {\bibinfo  {journal} {Phys.
  Rev. B}\ }\textbf {\bibinfo {volume} {95}},\ \bibinfo {pages} {035421}
  (\bibinfo {year} {2017})}\BibitemShut {NoStop}%
\bibitem [{\citenamefont {Rice}\ and\ \citenamefont
  {Mele}(1982)}]{PhysRevLett.49.1455}%
  \BibitemOpen
  \bibfield  {author} {\bibinfo {author} {\bibfnamefont {M.~J.}\ \bibnamefont
  {Rice}}\ and\ \bibinfo {author} {\bibfnamefont {E.~J.}\ \bibnamefont
  {Mele}},\ }\bibfield  {title} {\enquote {\bibinfo {title} {Elementary
  excitations of a linearly conjugated diatomic polymer},}\ }\href {\doibase
  10.1103/PhysRevLett.49.1455} {\bibfield  {journal} {\bibinfo  {journal}
  {Phys. Rev. Lett.}\ }\textbf {\bibinfo {volume} {49}},\ \bibinfo {pages}
  {1455--1459} (\bibinfo {year} {1982})}\BibitemShut {NoStop}%
\bibitem [{\citenamefont {Niu}(1990)}]{PhysRevLett.64.1812}%
  \BibitemOpen
  \bibfield  {author} {\bibinfo {author} {\bibfnamefont {Q.}~\bibnamefont
  {Niu}},\ }\bibfield  {title} {\enquote {\bibinfo {title} {Towards a quantum
  pump of electric charges},}\ }\href {\doibase 10.1103/PhysRevLett.64.1812}
  {\bibfield  {journal} {\bibinfo  {journal} {Phys. Rev. Lett.}\ }\textbf
  {\bibinfo {volume} {64}},\ \bibinfo {pages} {1812--1815} (\bibinfo {year}
  {1990})}\BibitemShut {NoStop}%
\bibitem [{\citenamefont {Berg}\ \emph {et~al.}(2011)\citenamefont {Berg},
  \citenamefont {Levin},\ and\ \citenamefont
  {Altman}}]{PhysRevLett.106.110405}%
  \BibitemOpen
  \bibfield  {author} {\bibinfo {author} {\bibfnamefont {E.}~\bibnamefont
  {Berg}}, \bibinfo {author} {\bibfnamefont {M.}~\bibnamefont {Levin}}, \ and\
  \bibinfo {author} {\bibfnamefont {E.}~\bibnamefont {Altman}},\ }\bibfield
  {title} {\enquote {\bibinfo {title} {Quantized pumping and topology of the
  phase diagram for a system of interacting bosons},}\ }\href {\doibase
  10.1103/PhysRevLett.106.110405} {\bibfield  {journal} {\bibinfo  {journal}
  {Phys. Rev. Lett.}\ }\textbf {\bibinfo {volume} {106}},\ \bibinfo {pages}
  {110405} (\bibinfo {year} {2011})}\BibitemShut {NoStop}%
\bibitem [{\citenamefont {Hayward}\ \emph {et~al.}(2018)\citenamefont
  {Hayward}, \citenamefont {Schweizer}, \citenamefont {Lohse}, \citenamefont
  {Aidelsburger},\ and\ \citenamefont {Heidrich-Meisner}}]{PhysRevB.98.245148}%
  \BibitemOpen
  \bibfield  {author} {\bibinfo {author} {\bibfnamefont {A.}~\bibnamefont
  {Hayward}}, \bibinfo {author} {\bibfnamefont {C.}~\bibnamefont {Schweizer}},
  \bibinfo {author} {\bibfnamefont {M.}~\bibnamefont {Lohse}}, \bibinfo
  {author} {\bibfnamefont {M.}~\bibnamefont {Aidelsburger}}, \ and\ \bibinfo
  {author} {\bibfnamefont {F.}~\bibnamefont {Heidrich-Meisner}},\ }\bibfield
  {title} {\enquote {\bibinfo {title} {Topological charge pumping in the
  interacting bosonic rice-mele model},}\ }\href {\doibase
  10.1103/PhysRevB.98.245148} {\bibfield  {journal} {\bibinfo  {journal} {Phys.
  Rev. B}\ }\textbf {\bibinfo {volume} {98}},\ \bibinfo {pages} {245148}
  (\bibinfo {year} {2018})}\BibitemShut {NoStop}%
\bibitem [{\citenamefont {Meidan}\ \emph {et~al.}(2011)\citenamefont {Meidan},
  \citenamefont {Micklitz},\ and\ \citenamefont
  {Brouwer}}]{PhysRevB.84.195410}%
  \BibitemOpen
  \bibfield  {author} {\bibinfo {author} {\bibfnamefont {D.}~\bibnamefont
  {Meidan}}, \bibinfo {author} {\bibfnamefont {T.}~\bibnamefont {Micklitz}}, \
  and\ \bibinfo {author} {\bibfnamefont {P.~W.}\ \bibnamefont {Brouwer}},\
  }\bibfield  {title} {\enquote {\bibinfo {title} {Topological classification
  of adiabatic processes},}\ }\href {\doibase 10.1103/PhysRevB.84.195410}
  {\bibfield  {journal} {\bibinfo  {journal} {Phys. Rev. B}\ }\textbf {\bibinfo
  {volume} {84}},\ \bibinfo {pages} {195410} (\bibinfo {year}
  {2011})}\BibitemShut {NoStop}%
\bibitem [{\citenamefont {Watanabe}\ and\ \citenamefont
  {Oshikawa}(2018)}]{PhysRevX.8.021065}%
  \BibitemOpen
  \bibfield  {author} {\bibinfo {author} {\bibfnamefont {Haruki}\ \bibnamefont
  {Watanabe}}\ and\ \bibinfo {author} {\bibfnamefont {Masaki}\ \bibnamefont
  {Oshikawa}},\ }\bibfield  {title} {\enquote {\bibinfo {title} {Inequivalent
  berry phases for the bulk polarization},}\ }\href {\doibase
  10.1103/PhysRevX.8.021065} {\bibfield  {journal} {\bibinfo  {journal} {Phys.
  Rev. X}\ }\textbf {\bibinfo {volume} {8}},\ \bibinfo {pages} {021065}
  (\bibinfo {year} {2018})}\BibitemShut {NoStop}%
\bibitem [{\citenamefont {King-Smith}\ and\ \citenamefont
  {Vanderbilt}(1993)}]{PhysRevB.47.1651}%
  \BibitemOpen
  \bibfield  {author} {\bibinfo {author} {\bibfnamefont {R.~D.}\ \bibnamefont
  {King-Smith}}\ and\ \bibinfo {author} {\bibfnamefont {D.}~\bibnamefont
  {Vanderbilt}},\ }\bibfield  {title} {\enquote {\bibinfo {title} {Theory of
  polarization of crystalline solids},}\ }\href {\doibase
  10.1103/PhysRevB.47.1651} {\bibfield  {journal} {\bibinfo  {journal} {Phys.
  Rev. B}\ }\textbf {\bibinfo {volume} {47}},\ \bibinfo {pages} {1651--1654}
  (\bibinfo {year} {1993})}\BibitemShut {NoStop}%
\bibitem [{\citenamefont {Wen}(2017)}]{RevModPhys.89.041004}%
  \BibitemOpen
  \bibfield  {author} {\bibinfo {author} {\bibfnamefont {X.-G.}\ \bibnamefont
  {Wen}},\ }\bibfield  {title} {\enquote {\bibinfo {title} {Colloquium: Zoo of
  quantum-topological phases of matter},}\ }\href {\doibase
  10.1103/RevModPhys.89.041004} {\bibfield  {journal} {\bibinfo  {journal}
  {Rev. Mod. Phys.}\ }\textbf {\bibinfo {volume} {89}},\ \bibinfo {pages}
  {041004} (\bibinfo {year} {2017})}\BibitemShut {NoStop}%
\bibitem [{\citenamefont {Anisimovas}\ \emph {et~al.}(2016)\citenamefont
  {Anisimovas}, \citenamefont {Ra\ifmmode \check{c}\else
  \v{c}\fi{}i\ifmmode~\bar{u}\else \={u}\fi{}nas}, \citenamefont {Str\"ater},
  \citenamefont {Eckardt}, \citenamefont {Spielman},\ and\ \citenamefont
  {Juzeli\ifmmode~\bar{u}\else \={u}\fi{}nas}}]{PhysRevA.94.063632}%
  \BibitemOpen
  \bibfield  {author} {\bibinfo {author} {\bibfnamefont {E.}~\bibnamefont
  {Anisimovas}}, \bibinfo {author} {\bibfnamefont {M.}~\bibnamefont {Ra\ifmmode
  \check{c}\else \v{c}\fi{}i\ifmmode~\bar{u}\else \={u}\fi{}nas}}, \bibinfo
  {author} {\bibfnamefont {C.}~\bibnamefont {Str\"ater}}, \bibinfo {author}
  {\bibfnamefont {A.}~\bibnamefont {Eckardt}}, \bibinfo {author} {\bibfnamefont
  {I.~B.}\ \bibnamefont {Spielman}}, \ and\ \bibinfo {author} {\bibfnamefont
  {G.}~\bibnamefont {Juzeli\ifmmode~\bar{u}\else \={u}\fi{}nas}},\ }\bibfield
  {title} {\enquote {\bibinfo {title} {Semisynthetic zigzag optical lattice for
  ultracold bosons},}\ }\href {\doibase 10.1103/PhysRevA.94.063632} {\bibfield
  {journal} {\bibinfo  {journal} {Phys. Rev. A}\ }\textbf {\bibinfo {volume}
  {94}},\ \bibinfo {pages} {063632} (\bibinfo {year} {2016})}\BibitemShut
  {NoStop}%
\bibitem [{\citenamefont {Greschner}\ \emph {et~al.}(2013)\citenamefont
  {Greschner}, \citenamefont {Santos},\ and\ \citenamefont
  {Vekua}}]{PhysRevA.87.033609}%
  \BibitemOpen
  \bibfield  {author} {\bibinfo {author} {\bibfnamefont {S.}~\bibnamefont
  {Greschner}}, \bibinfo {author} {\bibfnamefont {L.}~\bibnamefont {Santos}}, \
  and\ \bibinfo {author} {\bibfnamefont {T.}~\bibnamefont {Vekua}},\ }\bibfield
   {title} {\enquote {\bibinfo {title} {Ultracold bosons in zig-zag optical
  lattices},}\ }\href {\doibase 10.1103/PhysRevA.87.033609} {\bibfield
  {journal} {\bibinfo  {journal} {Phys. Rev. A}\ }\textbf {\bibinfo {volume}
  {87}},\ \bibinfo {pages} {033609} (\bibinfo {year} {2013})}\BibitemShut
  {NoStop}%
\bibitem [{\citenamefont {Chhajlany}\ \emph {et~al.}(2016)\citenamefont
  {Chhajlany}, \citenamefont {Grzybowski}, \citenamefont
  {Stasi\ifmmode~\acute{n}\else \'{n}\fi{}ska}, \citenamefont {Lewenstein},\
  and\ \citenamefont {Dutta}}]{PhysRevLett.116.225303}%
  \BibitemOpen
  \bibfield  {author} {\bibinfo {author} {\bibfnamefont {R.~W.}\ \bibnamefont
  {Chhajlany}}, \bibinfo {author} {\bibfnamefont {P.~R.}\ \bibnamefont
  {Grzybowski}}, \bibinfo {author} {\bibfnamefont {J.}~\bibnamefont
  {Stasi\ifmmode~\acute{n}\else \'{n}\fi{}ska}}, \bibinfo {author}
  {\bibfnamefont {M.}~\bibnamefont {Lewenstein}}, \ and\ \bibinfo {author}
  {\bibfnamefont {O.}~\bibnamefont {Dutta}},\ }\bibfield  {title} {\enquote
  {\bibinfo {title} {Hidden string order in a hole superconductor with extended
  correlated hopping},}\ }\href {\doibase 10.1103/PhysRevLett.116.225303}
  {\bibfield  {journal} {\bibinfo  {journal} {Phys. Rev. Lett.}\ }\textbf
  {\bibinfo {volume} {116}},\ \bibinfo {pages} {225303} (\bibinfo {year}
  {2016})}\BibitemShut {NoStop}%
\bibitem [{\citenamefont {Mandel}\ \emph
  {et~al.}(2003{\natexlab{b}})\citenamefont {Mandel}, \citenamefont {Greiner},
  \citenamefont {Widera}, \citenamefont {Rom}, \citenamefont {H{\"a}nsch},\
  and\ \citenamefont {Bloch}}]{mandel2003controlled}%
  \BibitemOpen
  \bibfield  {author} {\bibinfo {author} {\bibfnamefont {O.}~\bibnamefont
  {Mandel}}, \bibinfo {author} {\bibfnamefont {M.}~\bibnamefont {Greiner}},
  \bibinfo {author} {\bibfnamefont {A.}~\bibnamefont {Widera}}, \bibinfo
  {author} {\bibfnamefont {T.}~\bibnamefont {Rom}}, \bibinfo {author}
  {\bibfnamefont {T.}~\bibnamefont {H{\"a}nsch}}, \ and\ \bibinfo {author}
  {\bibfnamefont {I.}~\bibnamefont {Bloch}},\ }\bibfield  {title} {\enquote
  {\bibinfo {title} {Controlled collisions for multi-particle entanglement of
  optically trapped atoms},}\ }\href@noop {} {\bibfield  {journal} {\bibinfo
  {journal} {Nature}\ }\textbf {\bibinfo {volume} {425}},\ \bibinfo {pages}
  {937} (\bibinfo {year} {2003}{\natexlab{b}})}\BibitemShut {NoStop}%
\bibitem [{\citenamefont {Pachos}\ and\ \citenamefont
  {Knight}(2003)}]{PhysRevLett.91.107902}%
  \BibitemOpen
  \bibfield  {author} {\bibinfo {author} {\bibfnamefont {Jiannis~K.}\
  \bibnamefont {Pachos}}\ and\ \bibinfo {author} {\bibfnamefont {Peter~L.}\
  \bibnamefont {Knight}},\ }\bibfield  {title} {\enquote {\bibinfo {title}
  {Quantum computation with a one-dimensional optical lattice},}\ }\href
  {\doibase 10.1103/PhysRevLett.91.107902} {\bibfield  {journal} {\bibinfo
  {journal} {Phys. Rev. Lett.}\ }\textbf {\bibinfo {volume} {91}},\ \bibinfo
  {pages} {107902} (\bibinfo {year} {2003})}\BibitemShut {NoStop}%
\bibitem [{\citenamefont {Belmechri}\ \emph {et~al.}(2013)\citenamefont
  {Belmechri}, \citenamefont {Förster}, \citenamefont {Alt}, \citenamefont
  {Widera}, \citenamefont {Meschede},\ and\ \citenamefont
  {Alberti}}]{Belmechri_2013}%
  \BibitemOpen
  \bibfield  {author} {\bibinfo {author} {\bibfnamefont {N.}~\bibnamefont
  {Belmechri}}, \bibinfo {author} {\bibfnamefont {L.}~\bibnamefont {Förster}},
  \bibinfo {author} {\bibfnamefont {W.}~\bibnamefont {Alt}}, \bibinfo {author}
  {\bibfnamefont {A.}~\bibnamefont {Widera}}, \bibinfo {author} {\bibfnamefont
  {D.}~\bibnamefont {Meschede}}, \ and\ \bibinfo {author} {\bibfnamefont
  {A.}~\bibnamefont {Alberti}},\ }\bibfield  {title} {\enquote {\bibinfo
  {title} {Microwave control of atomic motional states in a spin-dependent
  optical lattice},}\ }\href {\doibase 10.1088/0953-4075/46/10/104006}
  {\bibfield  {journal} {\bibinfo  {journal} {J. Phys. B: At., Mol. Opt.
  Phys.}\ }\textbf {\bibinfo {volume} {46}},\ \bibinfo {pages} {104006}
  (\bibinfo {year} {2013})}\BibitemShut {NoStop}%
\bibitem [{\citenamefont {Liu}\ \emph {et~al.}(2004)\citenamefont {Liu},
  \citenamefont {Wilczek},\ and\ \citenamefont {Zoller}}]{PhysRevA.70.033603}%
  \BibitemOpen
  \bibfield  {author} {\bibinfo {author} {\bibfnamefont {W.~V.}\ \bibnamefont
  {Liu}}, \bibinfo {author} {\bibfnamefont {F.}~\bibnamefont {Wilczek}}, \ and\
  \bibinfo {author} {\bibfnamefont {P.}~\bibnamefont {Zoller}},\ }\bibfield
  {title} {\enquote {\bibinfo {title} {Spin-dependent hubbard model and a
  quantum phase transition in cold atoms},}\ }\href {\doibase
  10.1103/PhysRevA.70.033603} {\bibfield  {journal} {\bibinfo  {journal} {Phys.
  Rev. A}\ }\textbf {\bibinfo {volume} {70}},\ \bibinfo {pages} {033603}
  (\bibinfo {year} {2004})}\BibitemShut {NoStop}%
\bibitem [{\citenamefont {Krutitsky}(2016)}]{krutitsky2016ultracold}%
  \BibitemOpen
  \bibfield  {author} {\bibinfo {author} {\bibfnamefont {K.}~\bibnamefont
  {Krutitsky}},\ }\bibfield  {title} {\enquote {\bibinfo {title} {Ultracold
  bosons with short-range interaction in regular optical lattices},}\ }\href
  {\doibase 10.1016/j.physrep.2015.10.004} {\bibfield  {journal} {\bibinfo
  {journal} {Phys. Rep.}\ }\textbf {\bibinfo {volume} {607}},\ \bibinfo {pages}
  {1--101} (\bibinfo {year} {2016})}\BibitemShut {NoStop}%
\bibitem [{\citenamefont {Greiner}\ \emph {et~al.}(2002)\citenamefont
  {Greiner}, \citenamefont {Mandel}, \citenamefont {Esslinger}, \citenamefont
  {H{\"a}nsch},\ and\ \citenamefont {Bloch}}]{greiner2002quantum}%
  \BibitemOpen
  \bibfield  {author} {\bibinfo {author} {\bibfnamefont {M.}~\bibnamefont
  {Greiner}}, \bibinfo {author} {\bibfnamefont {O.}~\bibnamefont {Mandel}},
  \bibinfo {author} {\bibfnamefont {T.}~\bibnamefont {Esslinger}}, \bibinfo
  {author} {\bibfnamefont {T.~W.}\ \bibnamefont {H{\"a}nsch}}, \ and\ \bibinfo
  {author} {\bibfnamefont {I.}~\bibnamefont {Bloch}},\ }\bibfield  {title}
  {\enquote {\bibinfo {title} {Quantum phase transition from a superfluid to a
  mott insulator in a gas of ultracold atoms},}\ }\href {\doibase
  10.1038/415039a} {\bibfield  {journal} {\bibinfo  {journal} {Nature}\
  }\textbf {\bibinfo {volume} {415}},\ \bibinfo {pages} {39} (\bibinfo {year}
  {2002})}\BibitemShut {NoStop}%
\bibitem [{\citenamefont {Takahashi}(1977)}]{JPC1977Takahashi}%
  \BibitemOpen
  \bibfield  {author} {\bibinfo {author} {\bibfnamefont {M.}~\bibnamefont
  {Takahashi}},\ }\bibfield  {title} {\enquote {\bibinfo {title} {Half-filled
  hubbard model at low temperature},}\ }\href
  {http://stacks.iop.org/0022-3719/10/i=8/a=031} {\bibfield  {journal}
  {\bibinfo  {journal} {J. Phys. C: Solid State Phys.}\ }\textbf {\bibinfo
  {volume} {10}},\ \bibinfo {pages} {1289} (\bibinfo {year}
  {1977})}\BibitemShut {NoStop}%
\bibitem [{\citenamefont {Rhim}\ \emph {et~al.}(2018)\citenamefont {Rhim},
  \citenamefont {Bardarson},\ and\ \citenamefont
  {Slager}}]{PhysRevB.97.115143}%
  \BibitemOpen
  \bibfield  {author} {\bibinfo {author} {\bibfnamefont {J.-W.}\ \bibnamefont
  {Rhim}}, \bibinfo {author} {\bibfnamefont {J.~H.}\ \bibnamefont {Bardarson}},
  \ and\ \bibinfo {author} {\bibfnamefont {R.-J.}\ \bibnamefont {Slager}},\
  }\bibfield  {title} {\enquote {\bibinfo {title} {Unified bulk-boundary
  correspondence for band insulators},}\ }\href {\doibase
  10.1103/PhysRevB.97.115143} {\bibfield  {journal} {\bibinfo  {journal} {Phys.
  Rev. B}\ }\textbf {\bibinfo {volume} {97}},\ \bibinfo {pages} {115143}
  (\bibinfo {year} {2018})}\BibitemShut {NoStop}%
\end{thebibliography}%
\end{document}